\documentclass[sigconf,nonacm]{acmart}  

\renewcommand\footnotetextcopyrightpermission[1]{} 

\usepackage{balance}
\usepackage{booktabs}
\usepackage{graphicx}
\usepackage{subcaption}
\usepackage{makecell}
\usepackage{multirow}
\usepackage{xspace}
\usepackage{xurl}
\usepackage[english]{babel}
\usepackage[utf8]{inputenc}
\usepackage[linesnumbered,ruled,vlined]{algorithm2e}
\usepackage[noend]{algpseudocode}
\usepackage{threeparttable}
\usepackage{hhline}
\usepackage{framed}
\usepackage{listings} 
\usepackage{color}

\usepackage{amsmath}
\DeclareMathOperator*{\argmax}{argmax}

\SetKwInput{KwInput}{Input}                
\SetKwInput{KwOutput}{Output}              
\SetKwInput{KwResult}{Result}              

\lstdefinestyle{plainTextStyle}{
    language=,
    basicstyle=\ttfamily\tiny, 
    numbers=none,              
    showspaces=false,
    showstringspaces=false,
    frame=single,
    breaklines=true
}

\newcommand{\numversions}{97\xspace}
\newcommand{\uniqueclasses}{67\xspace}

\newcommand{\langversions}{15\xspace}

\newcommand{\allmediancovTot}{96.0}
\newcommand{\allsdcovTot}{5.7}

\newcommand{\allmediancovAdd}{96.7}
\newcommand{\allsdcovAdd}{6.3}

\newcommand{\allmedianLedrutextAPFD}{87.6}
\newcommand{\allsdLedrutextAPFD}{12.8}
\newcommand{\allmedianLedrubytecodeAPFD}{89.9}
\newcommand{\allsdLedrubytecodeAPFD}{12.9}
\newcommand{\allmedianLedrubytecodefilterAPFD}{92.1}
\newcommand{\allsdLedrubytecodefilterAPFD}{11.6}
\newcommand{\allmedianFASTtextAPFD}{85.4}
\newcommand{\allsdFASTtextAPFD}{20.5}
\newcommand{\allmedianFASTbytecodeAPFD}{93.2}
\newcommand{\allsdFASTbytecodeAPFD}{11.9}
\newcommand{\allmedianFASTbytecodefilterAPFD}{90.5}
\newcommand{\allsdFASTbytecodefilterAPFD}{14.8}

\newcommand{\allmediancovTotRealfault}{10.5}
\newcommand{\allsdcovTotRealfault}{378.6}

\newcommand{\allmediancovAddRealfault}{9.0}
\newcommand{\allsdcovAddRealfault}{589.2}

\newcommand{\allmedianLedrutextRealfault}{110.0}
\newcommand{\allsdLedrutextRealfault}{520.0}
\newcommand{\allmedianLedrubytecodeRealfault}{124.0}
\newcommand{\allsdLedrubytecodeRealfault}{565.0}
\newcommand{\allmedianLedrubytecodefilterRealfault}{78.0}
\newcommand{\allsdLedrubytecodefilterRealfault}{559.5}
\newcommand{\allmedianFASTtextRealfault}{125.0}
\newcommand{\allsdFASTtextRealfault}{890.0}
\newcommand{\allmedianFASTbytecodeRealfault}{32.0}
\newcommand{\allsdFASTbytecodeRealfault}{523.9}
\newcommand{\allmedianFASTbytecodefilterRealfault}{63.0}
\newcommand{\allsdFASTbytecodefilterRealfault}{740.4}

\newcommand{\clitextPreptime}{177780.8}

\newcommand{\clifastPreptime}{48.9}

\newcommand{\clifastbytecodePreptime}{8.1}

\newcommand{\clifastbytecodefilterPreptime}{2.7}

\newcommand{\clibytecodePreptime}{405.2}

\newcommand{\clibytecodefilterPreptime}{99.8}

\newcommand{\cliCovTotTotaltime}{35.5}
\newcommand{\cliCovAddTotaltime}{42.6}

\newcommand{\cliAvgLedruPtime}{1117.7}
\newcommand{\cliAvgFASTPtime}{9.8}

\newcommand{\compresstextPreptime}{505855.0}

\newcommand{\compressfastPreptime}{36.7}

\newcommand{\compressfastbytecodePreptime}{14.7}

\newcommand{\compressfastbytecodefilterPreptime}{5.4}

\newcommand{\compressbytecodePreptime}{1665.6}

\newcommand{\compressbytecodefilterPreptime}{664.1}

\newcommand{\compressCovTotTotaltime}{29.3}
\newcommand{\compressCovAddTotaltime}{37.6}

\newcommand{\compressAvgLedruPtime}{2792.6}
\newcommand{\compressAvgFASTPtime}{18.7}

\newcommand{\csvtextPreptime}{317075.3}

\newcommand{\csvfastPreptime}{49.5}

\newcommand{\csvfastbytecodePreptime}{16.6}

\newcommand{\csvfastbytecodefilterPreptime}{5.9}

\newcommand{\csvbytecodePreptime}{852.85}

\newcommand{\csvbytecodefilterPreptime}{250.95}

\newcommand{\csvCovTotTotaltime}{36.1}
\newcommand{\csvCovAddTotaltime}{46.7}

\newcommand{\csvAvgLedruPtime}{2145.5}
\newcommand{\csvAvgFASTPtime}{19.6}

\newcommand{\jsouptextPreptime}{1467823.5}

\newcommand{\jsoupfastPreptime}{80.4}

\newcommand{\jsoupfastbytecodePreptime}{25.3}

\newcommand{\jsoupfastbytecodefilterPreptime}{6.4}

\newcommand{\jsoupbytecodePreptime}{2931.1}

\newcommand{\jsoupbytecodefilterPreptime}{794.42}

\newcommand{\jsoupCovTotTotaltime}{446.1}
\newcommand{\jsoupCovAddTotaltime}{568.7}

\newcommand{\jsoupAvgLedruPtime}{9587.5}
\newcommand{\jsoupAvgFASTPtime}{43.7}

\newcommand{\langtextPreptime}{930756.6}

\newcommand{\langfastPreptime}{60.0}

\newcommand{\langfastbytecodePreptime}{21.9}

\newcommand{\langfastbytecodefilterPreptime}{7.3}

\newcommand{\langbytecodePreptime}{1142.6}

\newcommand{\langbytecodefilterPreptime}{332.6}

\newcommand{\langCovTotTotaltime}{279.4}
\newcommand{\langCovAddTotaltime}{344.1}

\newcommand{\langAvgLedruPtime}{5927.5}
\newcommand{\langAvgFASTPtime}{42.1}

\newcommand{\mathtextPreptime}{347037.9}

\newcommand{\mathfastPreptime}{88.9}

\newcommand{\mathfastbytecodePreptime}{33.2}

\newcommand{\mathfastbytecodefilterPreptime}{9.2}

\newcommand{\mathbytecodePreptime}{757.65}

\newcommand{\mathbytecodefilterPreptime}{261.4}

\newcommand{\mathCovTotTotaltime}{89.2}
\newcommand{\mathCovAddTotaltime}{118.5}

\newcommand{\mathAvgLedruPtime}{2818.7}
\newcommand{\mathAvgFASTPtime}{29.8}

\newcommand{\timetextPreptime}{633.3}

\newcommand{\timefastPreptime}{2.6}

\newcommand{\timefastbytecodePreptime}{0.9}

\newcommand{\timefastbytecodefilterPreptime}{0.7}

\newcommand{\timebytecodePreptime}{5.1}

\newcommand{\timebytecodefilterPreptime}{3.0}

\newcommand{\timeCovTotTotaltime}{2.2}
\newcommand{\timeCovAddTotaltime}{2.7}

\newcommand{\timeAvgLedruPtime}{3.7}
\newcommand{\timeAvgFASTPtime}{0.3}

\newcommand{\etal}{et~al.\xspace}
\newcommand{\ie}{i.e.\xspace}

\title[Bytecode Diversity for TCP]{Empirically Evaluating the Use of Bytecode for Diversity-Based Test Case Prioritisation}

\author{Islam T. Elgendy}
\thanks{This is the author's version submitted to arXiv. The final version will appear in the proceedings of EASE 2025.}

\affiliation{
  \institution{University of Sheffield}
  \country{Sheffield, UK}
}

\author{Robert M. Hierons}
\affiliation{
  \institution{University of Sheffield}
  \country{Sheffield, UK}
}

\author{Phil McMinn}
\affiliation{
  \institution{University of Sheffield}
  \country{Sheffield, UK}
}

\begin{document}

\begin{abstract}
Regression testing assures software correctness after changes but is resource-intensive. 
Test Case Prioritisation (TCP) mitigates this by ordering tests to maximise early fault detection. 
Diversity-based TCP prioritises dissimilar tests, assuming they exercise different system parts and uncover more faults. 
Traditional static diversity-based TCP approaches (\ie, methods that utilise the dissimilarity of tests), like the state-of-the-art FAST approach, rely on textual diversity from test source code, which is effective but inefficient due to its relative verbosity and redundancies affecting similarity calculations. 
This paper is the first to study bytecode as the basis of diversity in TCP, leveraging its compactness for improved efficiency and accuracy. 
An empirical study on seven Defects4J projects shows that bytecode diversity improves fault detection by 2.3–7.8\% over text-based TCP. 
It is also 2–3 orders of magnitude faster in one TCP approach and 2.5–6 times faster in FAST-based TCP. 
Filtering specific bytecode instructions improves efficiency up to fourfold while maintaining effectiveness, making bytecode diversity a superior static approach.     
\end{abstract}


\maketitle

\section{Introduction}
\label{sec:introduction}

Regression testing is a fundamental component of software testing. Its purpose is to verify the correctness of software after modifications to the software itself or its execution environment.
Although regression testing is important, it is widely recognised as a resource-intensive and time-consuming process~\cite{Yoo2010B}.
Test Case Prioritisation (TCP) is a primary strategy to mitigate this challenge.
TCP reorders test cases to maximise early fault detection, particularly when it has to be prematurely halted due to budgetary constraints~\cite{Yoo2010B}.
In the domain of software testing research, TCP has garnered significant attention~\cite{Lou2019, Khatibsyarbini2018, Henard2016, Yoo2010B}, with diversity-based testing techniques playing a pivotal role~\cite{Elgendy2025, Ledru2012, Miranda2018}.
Elgendy~\etal~\cite{Elgendy2025} provide a formal definition of diversity-based testing techniques, which quantify dissimilarity among test cases through a distance metric derived from various artefacts related to the test, such as inputs or the text of the test cases.
The idea of using test diversity is that dissimilar test cases are more likely to exercise different parts of the system than more similar test cases, increasing fault detection.

TCP approaches that utilise the source code of the test cases can be classified mainly as static techniques (\ie, techniques that analyse test cases' text), and dynamic techniques (\ie, techniques that require test suite execution).
An example of a static technique is using the source code of the test cases~\cite{Ledru2012}, while an example of a dynamic technique is using coverage information~\cite{Rothermel2001}.
Static approaches may be less effective, but they do not require any complex program instrumentation and program execution.

Many diversity-based TCP techniques leverage different testing artefacts, with static approaches that rely on test case text being commonly used and showing promising results~\cite{Ledru2012,Miranda2018,Wu2012,Wang2015,Altiero2024}.
These static techniques simplify the process by avoiding program instrumentation or model construction.
However, relying on test case text has inherent limitations.
The string-based representation of test cases is verbose and often large, increasing the time required for similarity calculations, particularly with complex distance metrics like Levenshtein distance~\cite{Elgendy2024}.
Furthermore, we hypothesise that test cases contain extraneous information for detecting faults such as variable names, comments, and assertion messages, which can distort similarity calculations and affect prioritisation.

Miranda~\etal~\cite{Miranda2018} introduced the FAST family of TCP approaches to address runtime concerns with large test suites by leveraging big data techniques such as Shingling, Minhashing, and Locality-Sensitive Hashing to compute test similarity for TCP.
However, these techniques still measure similarity based on test case text.
While preprocessing to remove comments and whitespace can mitigate some issues, it does not eliminate all irrelevant information.
A more concise and efficient representation of test cases could address these limitations by reducing the size of the data used for similarity calculations and eliminating irrelevant information.

In this paper, we propose to use the bytecode of the test cases as the basis for diversity calculations.
Bytecode, the intermediate code executed by a virtual machine, offers a more abstract representation than machine code but is more concrete than high-level programming language code.
The bytecode is smaller than the source code, as it is a condensed form of the original source code that strips away information such as comments, whitespace, and variable names.
It retains only execution-critical details, making it more efficient to store and process than source code.
While this may reduce diversity, our hypothesis it does not hinder effectiveness.
%
Moreover, not all bytecode instructions differentiate tests.
Some instructions, such as loading and storing variables in/from the stack, serve a mechanical role rather than providing insight into the logic of the test.
Therefore, we further propose to filter out these types of bytecode instructions and extract only instructions that relate to the functionality being tested, such as setting up specific values or calling methods that are central to the test's purpose.

We implemented two state-of-the-art static TCP approaches, Ledru-TCP~\cite{Ledru2012} and FAST-TCP~\cite{Miranda2018}, applying bytecode similarity for the first time instead of textual similarity.
We evaluated its effectiveness and efficiency against traditional static TCP approaches using textual diversity.
%
The Levenshtein distance measured textual similarity, as it suits string data representations~\cite{Elgendy2025} and varying test sizes, and the same metric was used for bytecode similarity to ensure consistency.
%
Results show that bytecode-based TCP outperforms text-based TCP, achieving 2.3\%–7.8\% higher APFD and up to 3.9 times better real-fault detection.
%
%
The filtering bytecode approach improved APFD by 4.5\% to 5.1\% and doubled real fault detection compared to textual information.

Additionally, bytecode diversity-based methods are significantly faster than textual diversity-based ones. 
Ledru-TCP using all bytecode information was 2–3 orders of magnitude faster than using text and 2.5 to 6 times faster in FAST-TCP.
%
Filtering bytecode information was 455–2798 times faster than text in Ledru-TCP and 4–17.9 times faster in FAST-TCP.

Finally, we evaluated how closely static bytecode diversity TCP matches dynamic TCP performance.
%
We implemented two dynamic TCP techniques using greedy total and greedy additional algorithms~\cite{Rothermel2001}.
As expected, dynamic approaches, with more information, outperformed bytecode diversity in fault detection, achieving 2.8\%–3.5\% higher APFDs.
However, the best bytecode TCP approach was 6.5–8.3 times faster and avoids the complexity of program instrumentation required for coverage-based techniques.
We provide a replication package~\cite{Replication2025} with detailed descriptions of the system.
The paper offers the following contributions:

\begin{enumerate}
\item The first known use of bytecode of test cases as a basis for measuring their diversity in a TCP technique.

\item A novel bytecode filtering method for further enhancing TCP efficiency.

\item An evaluation of bytecode diversity against textual diversity-based approaches, providing new insights into their effectiveness in mutant and real-fault detection, as well as runtime performance.

\item An empirical comparison of static and dynamic TCP approaches with bytecode diversity-based TCP, using data from seven projects and \numversions versions from Defects4J, offering insights into the benefits of bytecode diversity.
\end{enumerate}
\section{Background}
\label{sec:background}


In this section, we briefly explain TCP in general and diversity-based TCP approaches in particular.

\subsection{Test Case Prioritisation}
\label{sec:background-TCP}

\begin{figure}[t]
    \centering

    \includegraphics[width=\columnwidth]{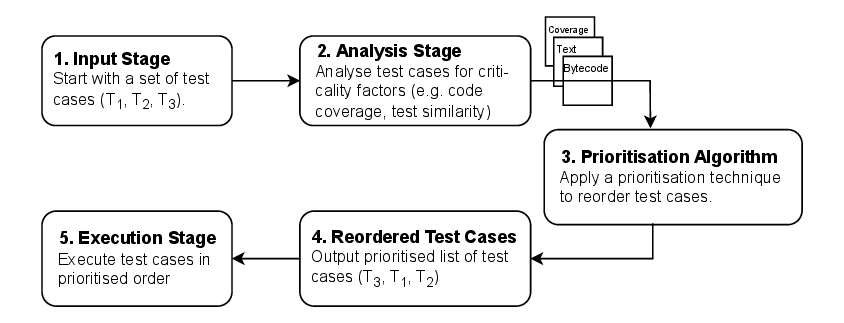} 
       
    \caption{TCP process diagram}
    \label{fig:TCP-process}
    \vspace{-1em}
\end{figure}

Test case prioritisation (TCP) re-orders test cases so that the tester gets the maximum benefit if testing is prematurely stopped at some arbitrary point due to some budget constraints~\cite{Yoo2010B}.
Figure~\ref{fig:TCP-process} presents the general TCP process diagram showing the main stages of TCP.
The primary goal of TCP is to identify and execute the most critical test cases earlier in the testing process.
This leads to earlier fault detection and better resource utilisation.
Rothermel~\etal~\cite{Rothermel2001} formally defined TCP as follows: Given a test suite~$T$, the set of permutations $P$ of $T$, and an award function $f$ from $P$ to real numbers, find ($T^\prime \in P$) such that $\forall T^{\prime\prime}$, where $T^{\prime\prime} \in P$ and $T^{\prime\prime} \neq T^{\prime}$, $f(T^{\prime}) \geq f(T^{\prime\prime})$.
The award function $f$ can be based on several criteria, such as code coverage~\cite{Rothermel1999,Rothermel2001}, fault detection history~\cite{Kim2002,Haghighatkhah2018B}, high-risk features~\cite{Elbaum2001}, software requirements~\cite{Srikanth2005,Arafeen2013}, or diversity of test cases~\cite{Ledru2012,Miranda2018}.

A well-known metric to evaluate TCP approaches is the Average Percentage of Faults Detected (APFD).
The faults can be real faults or mutants.
It is calculated using the following equation:

\begin{equation}
   \label{eq:APFD}
   \mathit{APFD} = 1 - \frac{\mathit{TF_1} + \mathit{TF_2} + \dots + \mathit{TF_m}}{nm} + \frac{1}{2n}
\end{equation}
\noindent where $n$ is the number of tests in the test suite, $m$ is the number of killable faults, and $\mathit{TF_{i}}$ is the position of the first test case that detects the fault $i$.

\subsection{Diversity-Based TCP}
\label{sec:background-diversity-TCP}

In this work, we focus on diversity-based TCP, where the function $f$ assesses test cases based on their diversity, either by rewarding those that are most dissimilar to others in an incrementally constructed subset $P$ or by evaluating the overall diversity of the entire test suite.
%
%
Elgendy~\etal~\cite{Elgendy2025} provided a recent survey on diversity-based testing techniques in software testing in general and covered diversity-based techniques in TCP.
First, diversity-based TCP collects the relevant information of the test case.
These can be the source code of the test cases, the inputs to the test, the outputs of the test cases, etc.
Then, the pairwise dissimilarity values between test cases are measured according to some distance metric, such as Euclidean distance, Jaccard distance, Levenshtein distance, etc.
A ``dissimilarity value'' is a measure that quantifies how different two test cases are, typically based on shared characteristics or features.
In our work, a lower dissimilarity value indicates greater similarity, while a higher value reflects greater dissimilarity.
Finally, test cases are ordered based on the dissimilarity values.
The ordering occurs in iterations, where in each iteration a test case is removed from the test suite and added to the new permutation set.
The first test case to be added to the permuted set can be selected randomly, or might be the most dissimilar test case from all others.
Then, subsequent tests are selected one at a time, where the most dissimilar test from the ones already in the permuted set is selected, and the process continues until the test suite is completed.

Ledru~\etal~\cite{Ledru2012} proposed such a diversity-based TCP approach.
They used five different distance metrics to calculate the pairwise dissimilarity values between test cases.
The dissimilarity values are stored in a ``similarity matrix''.
Miranda~\etal~\cite{Miranda2018} developed the FAST family of TCP techniques, which are quick and scalable for large test suites.
They applied some big data techniques, such as Shingling, Minhashing, and Locality-Sensitive Hashing to calculate the dissimilarity between test cases.
In both works, the source code of the test cases is used as a basis to calculate dissimilarity.
We will expand more on these works in Section~\ref{sec:TCP-approaches}.

\subsection{Bytecode}
\label{sec:background-bytecode}

Bytecode is an intermediate representation of source code, designed to be executed by a virtual machine rather than directly by the hardware.
Bytecode is generated by a compiler from the source code and is typically more compact, as it omits non-essential elements like comments and variable names.
This compactness enhances processing efficiency and reduces storage requirements, making bytecode an attractive artefact for TCP.

Java Bytecode is the Java Virtual Machine (JVM)-executable intermediate representation compiled from Java source code. 
%
%
%
Java Bytecode consists of a set of instructions and operands encoded in a binary format, providing a structured and efficient means of representing the program's logic and structure.
ASM~\cite{Asm2025}, a widely used library, enables dynamic bytecode manipulation, supporting class transformation, instrumentation, and analysis.

\section{Approach}
\label{sec:approach}

In this section, we show a motivating example and describe the TCP approaches used in our study.

\subsection{Motivating Example}

Consider three test cases written by developers for the Apache Commons Math project shown in Figure~\ref{fig:example} parts a)-c).
We made a variation of test case 3, to showcase the problem, shown in Figure~\ref{fig:example-4}.
The test cases are designed to test the \texttt{BigFraction} class.
Figure~\ref{fig:example-bytecode} shows the bytecodes of the test cases shown in Figure~\ref{fig:example} in the human-readable and hexadecimal forms.

Let us take the example of the method \texttt{testFormatNegative} shown in Figure~\ref{fig:example-2} and its corresponding bytecode shown in the second column of Figure~\ref{fig:example-bytecode1} to illustrate how high-level Java instructions translate into low-level bytecode instructions.
In the source code, a \texttt{BigFraction} object is created and initialised with the values -1 and 2.
The bytecode reflects this sequence starting with the \texttt{new} instruction at index 0, which allocates memory for the new \texttt{BigFraction} object.
The \texttt{dup} instruction at index 3 duplicates the top value of the stack to prepare for the constructor invocation, and the \texttt{iconst\_m1} and \texttt{iconst\_2} instructions at 4 and 5 push the values -1 and 2 onto the stack, respectively.
The \texttt{invokespecial} instruction at 6 calls the constructor to initialise the object.
Bytecode method invocation instructions are represented by \texttt{invokevirtual}, \texttt{invokestatic}, and \texttt{invokespecial}.
Other bytecode instructions that involve loading and storing references to/from local variables are \texttt{aload} and \texttt{astore}, respectively.
Field access instructions, like \texttt{properFormat} or \texttt{improperFormat.format}, are represented by \texttt{getfield}, \texttt{putfield}, and \texttt{getstatic}.
%

\begin{figure}[t]
    \centering
    \begin{subfigure}[b]{0.45\textwidth}
        \centering
        \begin{lstlisting}
public void testParseBig() {
    BigFraction f1 = improperFormat.parse("16721..." +
                " / " +  "532255...");
    Assert.assertEquals(FastMath.PI, f1.doubleValue(), 0.0);
    BigFraction f2 = properFormat.parse("3 " +
                "75363..." + " / " + "53225...");
    Assert.assertEquals(FastMath.PI, f2.doubleValue(), 0.0);
    Assert.assertEquals(f1, f2);
    BigDecimal pi = new BigDecimal("3.14159...");
    Assert.assertEquals(pi, f1.bigDecimalValue(99, BigDecimal.ROUND_HALF_EVEN));
}
        \end{lstlisting}
        \vspace{-1em}
    \caption{Test case 1}
    \label{fig:example-1}
    \end{subfigure}
    
    \hfill

    \begin{subfigure}[b]{0.45\textwidth}
        \centering
        \begin{lstlisting}[label=lst:testFormatNegative]
public void testFormatNegative() {
    BigFraction c = new BigFraction(-1, 2);
    String expected = "-1 / 2";

    String actual = properFormat.format(c);
    Assert.assertEquals(expected, actual);

    actual = improperFormat.format(c);
    Assert.assertEquals(expected, actual);
}
        \end{lstlisting}
        \vspace{-1em}
    \caption{Test case 2}
    \label{fig:example-2}
    \end{subfigure}
    
    \hfill

    \begin{subfigure}[b]{0.45\textwidth}
        \centering
        \begin{lstlisting}[label=lst:testFormatZero]
public void testFormatZero() {
    BigFraction c = new BigFraction(0, 1);
    String expected = "0 / 1";

    String actual = properFormat.format(c);
    Assert.assertEquals(expected, actual);

    actual = improperFormat.format(c);
    Assert.assertEquals(expected, actual);
}
        \end{lstlisting}
        \vspace{-1em}
    \caption{Test case 3 (original)}
    \label{fig:example-3}
    \end{subfigure}

    \hfill

\begin{subfigure}[b]{0.45\textwidth}
    \centering
    \begin{lstlisting}[label=lst:testFormatZero]
public void testFormatZero() {
    BigFraction bigFraction = new BigFraction(0, 1);
    // The expected output
    String output = "0 / 1";

	// The actual output
    String actual = properFormat.format(bigFraction);
    Assert.assertEquals(output, actual);

    actual = improperFormat.format(bigFraction);
    Assert.assertEquals(output, actual);
}
        \end{lstlisting}
        \vspace{-1em}
    \caption{Test case 3 (variation)}
    \label{fig:example-4}
    \end{subfigure}
    \vspace{-1em}
    \caption{An example of four test cases from the Math project, with test case 3 having two variations}
    \label{fig:example}
    \vspace{-0.2cm}
\end{figure}

The textual test cases from Figure~\ref{fig:example} are 705, 258, 252, and 320 bytes, respectively.
Textual representations of test cases include additional elements such as variable names, comments, and formatting, which do not contribute to test behaviour but can inflate dissimilarity calculations. 
For instance, \texttt{testFormatZero} has two variations (Figures~\ref{fig:example-3} and~\ref{fig:example-4}), differing only in variable names and comments—purely informational elements to the tester that do not affect execution. 
However, textual diversity methods treat these variations as more dissimilar than they actually are. 
In contrast, bytecode diversity captures only the core executable logic, leading to a more compact and precise representation. 
The two variations of \texttt{testFormatZero} produce identical bytecode despite textual differences, as shown in the last column in Figure~\ref{fig:example-bytecode}.
%

%
The bytecode sizes are just 48, 37, and 37 bytes, respectively.
The significant difference in size between the textual representation and the bytecode form highlights the potential inefficiencies of using textual diversity in TCP.
Using Levenshtein distance, the textual dissimilarity between test cases 1 and 2 is 545, whereas their bytecode dissimilarity is only 31. 
For test case 2 (Figure~\ref{fig:example-2}) and the original test case 3 (Figure~\ref{fig:example-3}), the textual dissimilarity is 13, increasing to 69 in the second variation (Figure~\ref{fig:example-4}) due to extraneous information (\ie, a difference of 56). 
In contrast, both variations of test case 3 produce identical bytecode, resulting in a consistent bytecode dissimilarity of just 2.

Clearly, bytecode is significantly smaller than its textual counterpart, reducing the computational overhead involved in calculating similarity values.
This can lead to faster execution of TCP processes.
Also, by focusing on the actual executable logic, bytecode diversity can avoid being potentially misled by superficial differences in the textual representation, leading to a more accurate assessment of the diversity between test cases.

\begin{figure}[t]
    \centering
    \begin{subfigure}[b]{0.45\textwidth}
        \centering
        \lstset{style=plainTextStyle}
        \begin{lstlisting}
testParseBig:			testFormatNegative:		testFormatZero:
0: aload_0				0: new            #8	0: new            #8
1: getfield       #3	3: dup					3: dup
4: ldc            #50	4: iconst_m1			4: iconst_0
6: invokevirtual  #19	5: iconst_2				5: iconst_1
9: astore_1				6: invokespecial  #9	6: invokespecial  #9
10: ldc2_w        #52	9: astore_1				9: astore_1
13: aload_1				10: ldc           #13	10: ldc           #14
14: invokevirtual #54	12: astore_2			12: astore_2
17: dconst_0			13: aload_0				13: aload_0
18: invokestatic  #55	14: getfield      #2	14: getfield      #2
21: aload_0				17: aload_1				17: aload_1
22: getfield      #2	18: invokevirtual #11	18: invokevirtual #11
25: ldc           #56	21: astore_3			21: astore_3
27: invokevirtual #19	22: aload_2				22: aload_2
30: astore_2			23: aload_3				23: aload_3
31: ldc2_w        #52	24: invokestatic  #12	24: invokestatic  #12
34: aload_2				27: aload_0				27: aload_0
35: invokevirtual #54	28: getfield      #3	28: getfield      #3
38: dconst_0			31: aload_1				31: aload_1
39: invokestatic  #55	32: invokevirtual #11	32: invokevirtual #11
42: aload_1				35: astore_3			35: astore_3
43: aload_2				36: aload_2				36: aload_2
44: invokestatic  #12	37: aload_3				37: aload_3
47: new           #57	38: invokestatic  #12	38: invokestatic  #12
50: dup					41: return				41: return
51: ldc           #58
53: invokespecial #59
56: astore_3
57: aload_3
58: aload_1
59: bipush        99
61: bipush        6
63: invokevirtual #60
66: invokestatic  #12
69: return
        \end{lstlisting}
        \vspace{-1em}
    \caption{The human-readable form}
    \label{fig:example-bytecode1}
    \end{subfigure}
    
    \hfill

    \begin{subfigure}[b]{0.45\textwidth}
        \centering
        \lstset{style=plainTextStyle}
        \begin{lstlisting}
testParseBig: 		
19 00 B4 12 B6 3A 01 12 19 01 B6 0E B8 19 00 B4 12 B6 3A 02 12 19 02 B6 0E B8 19 01 19 02 B8 BB 59 12 B7 3A 03 19 03 19 01 10 63 10 06 B6 B8 B1

testParseBig-filtered:
B4 B6 B6 0E B8 B4 B6 B6 0E B8 B8 59 B7 10 63 10 06 B6 B8 B1 

testFormatNegative: 
BB 59 02 05 B7 3A 01 12 3A 02 19 00 B4 19 01 B6 3A 03 19 02 19 03 B8 19 00 B4 19 01 B6 3A 03 19 02 19 03 B8 B1

testFormatNegative-filtered: 
59 02 05 B7 B4 B6 B8 B4 B6 B8 B1 

testFormatZero: 	
BB 59 03 04 B7 3A 01 12 3A 02 19 00 B4 19 01 B6 3A 03 19 02 19 03 B8 19 00 B4 19 01 B6 3A 03 19 02 19 03 B8 B1 

testFormatZero-filtered: 
59 03 04 B7 B4 B6 B8 B4 B6 B8 B1
        \end{lstlisting}
        \vspace{-1em}
    \caption{The Hexadecimal form}
    \label{fig:example-bytecode2}
    \end{subfigure}

    \caption{The human-readable and the Hexadecimal formats of the bytecodes for the tests in Figure~\ref{fig:example}}
    \label{fig:example-bytecode}
    \vspace{-0.5cm}
\end{figure}

\subsection{Filtering Bytecode}
When analysing the diversity of test cases based on their bytecode instructions, note that not all instructions contribute equally to the logical complexity or uniqueness of a test. 
Certain bytecode instructions, while essential for execution, are ubiquitous and serve a mechanical role rather than reflecting the test's unique logic. 
For instance, instructions like \texttt{aload} (loading a variable onto the stack) and \texttt{astore} (storing a variable from the stack) are necessary for code execution but are common across most tests and do not provide meaningful insight into the test's distinct behaviour.

In contrast, instructions such as \texttt{iconst} (pushing constant values onto the stack) or method invocations like \texttt{invokevirtual} and \texttt{invokestatic} better reflect the test's intent and logic. 
These instructions directly relate to the functionality being tested, such as initialising specific values or invoking methods that are central to the test's purpose. 
By focusing on these substantive instructions, diversity calculations can more accurately capture the variation in test logic and functionality by excluding routine operations that do not contribute to the core logic of the test case.

Filtering out routine bytecode instructions from diversity calculations will have a significant advantage: it reduces the overall size of the resulting bytecode.
Many bytecode instructions, such as \texttt{aload} and \texttt{astore}, occur frequently and contribute to the bulk of the bytecode without adding meaningful diversity in terms of test logic.
By excluding these routine instructions, the size of the bytecode representation for each test case becomes smaller and more concise.
Figure~\ref{fig:example-bytecode2} shows the hexadecimal forms of the bytecode and the filtered versions.

This filtering approach allows for a more meaningful analysis of the test suite's variety, highlighting the differences in test goals and behaviours rather than superficial variations in how variables are manipulated.
Also, the reduction in bytecode size has a direct impact on the efficiency of diversity calculations.
With a smaller set of instructions to analyse, the computational overhead for comparing test cases decreases significantly.
This leads to faster processing times, making it feasible to evaluate large test suites more efficiently.

\subsection{TCP Approaches}
\label{sec:TCP-approaches}

We implemented Ledru-TCP~\cite{Ledru2012} and FAST-TCP~\cite{Miranda2018}, two state-of-the-art TCP approaches.
Miranda~\etal~\cite{Miranda2018} proposed a family of TCP approaches, and we used one of these, FAST-pw.
We chose to use FAST-pw because they reported it to be the most precise in the ranking, as it guarantees the selection of the most dissimilar test case from the ones already prioritised.
We will simply refer to it as FAST from now on.
Ledru-TCP and FAST-TCP are general TCP algorithms that can operate on either textual or bytecode information. 
When using textual information, they are referred to as Ledru-Text and FAST-Text, respectively, and when using bytecode, they are referred to as Ledru-Bytecode and FAST-Bytecode.

\begin{algorithm}[!t]
    \DontPrintSemicolon
      
      \KwInput{List of tests $T$; Similarity matrix $Sim_{T}$}
      \KwOutput{Prioritised test suite $P$}

      $P$ $\leftarrow$ $\emptyset$ \\
      \tcp{First selection}
      $\mathit{minValues}$ $\leftarrow$ $\mathit{getMinValues}$($Sim_{T}$, $T$) \\
      $\mathit{maxKey}$ = $\mathit{getMaxOfMins}$($\mathit{minValues}$)  \\
      $P$ $\leftarrow$ $\mathit{Append}$($P$, $T[\mathit{maxKey}]$) \\
      $T$ $\leftarrow$ $\mathit{Remove}$($T$, $T[\mathit{maxKey}]$) \\
      \tcp{Subsequent selections}
      \While{$|T|$ $\neq$ $0$}
      {
        $\mathit{minValues}$ $\leftarrow$ $\mathit{getMinValues}$($Sim_{T}$, $P$) \\
        $\mathit{maxKey}$ = $\mathit{getMaxOfMins}$($\mathit{minValues}$)  \\
        $P$ $\leftarrow$ $\mathit{Append}$($P$, $T[\mathit{maxKey}]$) \\
        $T$ $\leftarrow$ $\mathit{Remove}$($T$, $T[\mathit{maxKey}]$) \\
      }
      return $P$
    
       \caption{Ledru-TCP algorithm, summarised from~\cite{Ledru2012}}
       \label{algorithm:selection}
\end{algorithm}

For Ledru-Bytecode, we used the ASM Java library to read the bytecode of the test cases and applied the Levenshtein distance~\cite{Levenshtein1966} to calculate the pairwise dissimilarity values and save them in a similarity matrix.
Algorithm~\ref{algorithm:selection} shows the steps of the Ledru-TCP~\cite{Ledru2012}.
%
The information we have is the pairwise similarity matrix $Sim_{T}$, and the list of tests $T$.
The similarity matrix is a two-dimensional array, where each cell contains the textual/bytecode dissimilarity value between two test cases.
We keep another list of the permuted tests $P$, which is initially empty (Line~1).
First, using the similarity matrix, the minimum value of each row is saved into an array $\texttt{minValues}$ (Line~2).
For the first selection, we get the test that has the maximum value of the $\texttt{minValues}$ (Line~3).
This is the test that is the farthest away from all the rest.
If multiple candidates have the same maximum value, we pick the first one.
We add this test into $P$ and remove it from $T$ (Lines~4 and 5).

\looseness=-1
For the subsequent selections, we calculate the minimum values of $P$ from the matrix and save them into $\texttt{minValues}$ (Line~7).
Then we get the test that has the maximum value of the new $\texttt{minValues}$ array (Line~8).
This is the test that is the farthest away from all the tests already selected in P.
This test is added to $P$ and removed from $T$ (Lines~9 and 10), repeating until all tests are selected.
%

\looseness=-1
In FAST-Bytecode, we used the ASM Java library to read the bytecodes, which are then converted into a hexadecimal format to serve as the encoded representation of the test suite.
Algorithm~\ref{algorithm:FAST} outlines the steps involved in FAST-TCP~\cite{Miranda2018}.
Initially, the permuted test suite $P$ is empty (Line~1), and the indices of the test cases are stored in $I$ (Line 2).
Using the encoded representation (text or bytecode) of the test suite $T$, FAST generates the MinHash signatures $M$ (Line~3).
The algorithm then computes the Locality-Sensitive Hashing (LSH) buckets $B$ (Line~4), while the cumulative signatures of the ordered test cases, denoted as $M(v)$, are initially set to empty (Line~5).

The candidate sets are generated within a while loop (Lines~6-15), where the prioritisation process occurs.
$M(v)$ is divided into $b$ bands, each of which is hashed.
If a collision is detected with the corresponding bucket in $B$, the test cases within that bucket are added to the candidate set $C_s$ (Line~7).
The actual candidate set $C_d$, used by FAST, is calculated as the complement of $C_s$, excluding the test cases that have already been prioritised in $P$ (Line~11).
FAST then selects the candidate test case that is farthest from $M(v)$ based on the Jaccard distance (Line~12).
$M(v)$ is updated to reflect the cumulative signatures of $P$ (Line~13).
Lastly, the selected test case is removed from $M$ and added to $P$ (Lines 14 and 15).

\begin{algorithm}[!t]
    \DontPrintSemicolon
      
      \KwInput{Coded test suite $T$}
      \KwOutput{Prioritised test suite $P$}

      $P$ $\leftarrow$ $\emptyset$ \\
      $I$ $\leftarrow$ $\mathit{GetTextCaseIDs}(T)$ \\
      $M$ $\leftarrow$ $\mathit{MHSignatures}(T)$ \\
      $B$ $\leftarrow$ $\mathit{LSHBuckets}(M)$ \\
      $M(v)$ $\leftarrow$ $\mathit{MHSignatures}(\emptyset)$ \\
      
      \While{$|P|$ $\neq$ $|I|$}
      {
        $C_{s}$ $\leftarrow$ $\mathit{LSHCandidates}(B, M(v))$ \\
        \If{$C_{s} = \emptyset$}
        {
          $M(v)$ $\leftarrow$ $\mathit{MHSignatures}(\emptyset)$ \\
          $C_{s}$ $\leftarrow$ $\mathit{LSHCandidates}(B, M(v))$ \\
        }
        $C_{d}$ $\leftarrow$ $(I - P - C_{s})$ \\
        $s$ $\leftarrow$ $\argmax_{c \in C} \{\mathit{JaccardDistance}(M(v), M(c)) \}$ \\
        $M(v)$ $\leftarrow$ $\mathit{UpdateMHSignatures}(M(v), M, s)$ \\
        $M$ $\leftarrow$ $\mathit{Remove}$($M$, $s$) \\
        $P$ $\leftarrow$ $\mathit{Append}$($P$, $s$) \\
      }
      return $P$
    
       \caption{FAST-TCP algorithm~\cite{Miranda2018}}
       \label{algorithm:FAST}    
\end{algorithm}
\section{Evaluation}
\label{sec:evaluation}

\begin{table*}[htbp]
    \centering

        \caption{The test subjects used in our work }
        
        \begin{tabular}{@{}l@{~}lr@{~~}r@{~~}r@{~~~~}r@{~~~~}r@{~~~~}r@{~~~~}r@{~~~~}r@{~~~~}}
            \toprule
            {\bf ID} & {\bf Project name} & {\bf \#Versions} & \multirow{2}{*}{\bf \makecell{ \#Unique \\ Classes}} & \multirow{2}{*}{\bf \makecell{ \#Real \\ Faults}} & {\bf \#Mutants} 
            & \multicolumn{4}{c}{\bf \# Test Cases}
            \\
            \cmidrule(lr){7-10}
                 & & & & & & {\bf Randoop } & {\bf EvoSuite } & {\bf Developer} & {\bf Total}
                \\
    
            \midrule
    
            Cli & jfreechart & 5 & 4 & 7 & 851 & 7425 & 279 & 242 & 7946 \\ 
Compress & commons-compress & 7 & 5 & 7 & 1051 & 10769 & 133 & 38 & 10940 \\ 
Csv & commons-csv & 10 & 5 & 10 & 767 & 15050 & 260 & 550 & 15860 \\ 
Jsoup & jsoup & 16 & 10 & 16 & 767 & 20575 & 278 & 632 & 21485 \\ 
Lang & commons-lang & \langversions & 11 & \langversions & 4136 & 25893 & 1103 & 688 & 27684 \\ 
Math & commons-math & 42 & 36 & 50 & 10986 & 27235 & 962 & 1612 & 29809 \\ 
Time & joda-time & 2 & 2 & 2 & 268 & 499 & 172 & 69 & 740 \\ 
\hline
Total & & \numversions & \uniqueclasses & 107 & 18826 & 107446 & 3187 & 3831 & 114464 \\
    
            \bottomrule
        \end{tabular}
        \label{table:subjects}
  
    \vspace{-0.2cm}
\end{table*}

We evaluated the proposed use of bytecode as a diversity artefact in TCP with the textual artefact in two static TCP approaches.
We described the diversity-based TCP approaches in Section~\ref{sec:TCP-approaches}.

Additionally, we compared the static TCP approach using bytecode with dynamic TCP approaches.
Thus, we implemented two coverage-based TCP techniques, Greedy Total and Greedy Additional.
First, we selected the test case with the highest coverage.
If two tests have the same coverage, the first one is selected.
Then we selected the next test cases based on the total maximum coverage, in the case of greedy total, or based on the maximum additional coverage, in the case of greedy additional.
If there is more than one possible test case with maximum coverage, the first one is selected.
For greedy additional, when max coverage is reached, we used a stacking approach, where coverage information is reset and the selection process continues as before.
This process continued until the last test case was selected.

\subsection{Research Questions}
\label{sec:RQs}

This study introduces the use of bytecode diversity as a diversity artefact in diversity-based TCP techniques.
We answer the following research questions:

\textbf{RQ1: Bytecode vs. Textual Diversity}
How does bytecode diversity compare to textual diversity in the context of diversity-based TCP techniques, such as those proposed by Ledru and FAST?
This RQ seeks to evaluate the effectiveness and applicability of bytecode diversity as an alternative to textual diversity.

\textbf{RQ2: Filtering Bytecode}
What is the impact of filtering bytecode instructions on the effectiveness and efficiency of static TCP approaches?
RQ2 explores whether reducing the set of bytecode instructions considered can improve performance (speed) while maintaining or enhancing prioritisation quality.

\textbf{RQ3: Bytecode TCP vs. Dynamic TCP}
How does bytecode diversity, as a static TCP technique, perform compared to dynamic techniques that rely on coverage information?
Finally, RQ3 aims to assess how close the performance of a static TCP approach based on bytecode is to that of dynamic TCP approaches.



\subsection{Evaluation Metrics}
\label{sec:eval-metrics}

To assess the effectiveness and efficiency of testing processes in our study, we used the following evaluation metrics.
%
We used mutation testing evaluated by APFD defined in Equation~\ref{eq:APFD} and described in Section~\ref{sec:background-TCP}.
Also, we utilised the real faults recorded in Defects4J~\cite{Just2014A} and used the number of the first test to reveal the real fault as the evaluation metric.
A test that can reveal a real fault is one that fails in the buggy version of the Defects4J but passes in the fixed version.

In order to evaluate the efficiency of the TCP approaches, we recorded the wall clock time of each TCP approach.
For diversity-based techniques, there is a stage to calculate the similarity matrix for Ledru-TCP, or the signatures for FAST-TCP.
This stage is referred to as {\it ``preparation time''}.
Then, there is the {\it ``prioritisation time''}, which is the time taken to reorder the test suite based on the various TCP approaches.
%
%
Obtaining coverage information requires either instrumenting the program under test and running the entire test suite or executing each test case individually to record coverage data. 
In this work, we adopted the latter approach.
However, to prevent unfair treatment of coverage-based techniques, as instrumentation would be considerably faster, we excluded preparation times and concentrated solely on prioritisation times.

\vspace{-0.2cm}
\subsection{Subjects}
\label{sec:subjects}

We ran our experiments using Java projects from the Defects4J framework~\cite{Just2014A}.
%
While we initially considered all available projects, we selected seven Maven-based projects where JaCoCo~\cite{JaCoCo2024} successfully generated coverage reports and PIT mutation analysis~\cite{Coles2016} ran without issues. 
Other projects were excluded due to failures in meeting these requirements.
We attempted to debug these issues for a week but ultimately did not pursue them further.

For our test suites, we used the developer-written tests augmented with automatically generated tests from Randoop~\cite{Pacheco2007} and EvoSuite~\cite{Fraser2011}.
We undertook this to create large, diverse test pools that possess high fault detection rates (\ie, mutation score) and can identify the real faults.
Table~\ref{table:subjects} shows the projects and the size of the test suites for each project used in our study, where a total of \numversions versions and \uniqueclasses unique classes are used in our experiments.
Also, we reported the number of mutants generated using the PIT mutation tool, the number of Randoop tests, the number of EvoSuite tests, and the number of developer-written tests for each project.
PIT is a widely adopted tool in both academic and industrial settings, and we employed Randoop and EvoSuite, which are state-of-the-art tools for automated test generation in Java.

In Defects4J, the targeted classes of two different versions can be the same, but the real fault can be different.
For example, the project Math version 26 (\texttt{Math26}) and version 27 (\texttt{Math27}) both target the ``\texttt{Fraction}'' class.
The fault in \texttt{Math26} is in the constructor of the class, while the fault in \texttt{Math27} is in the method ``\texttt{percentageValue}''.
As a result, the number of unique classes is lower than the number of versions.

%
We configured PIT to target the classes specified by Defects4J as target classes for which there are real faults.
For example, in Math26, we configured PIT to target the ``Fraction'' class, since the real fault is in the ``Fraction'' class.

\subsection{Methodology}
\label{sec:methodology}

In this section, we describe our methodology for obtaining the textual and bytecode information, mutation data, real fault detection, runtime, and coverage information, all of which are to be used in the static and dynamic TCP approaches.
We developed a tool to automate all the steps for our experiments.

\subsubsection{Data Collection}

Our tool parses the test files' source code and extracts each test case.
The lines of each test case are concatenated into a single string, where the tool calculates the pairwise dissimilarity values between the strings of test cases using Levenshtein distance.
We used the Levenshtein distance because it was reported by Elgendy~\etal~\cite{Elgendy2025} to be more appropriate with string data, which is the nature of our test cases.
In another study, Elgendy~\etal~\cite{Elgendy2024} reported that Levenshtein distance performs well as a string distance metric.
Then, a textual similarity CSV file is generated that represents the textual similarity matrix.
We compiled the source code using Apache Maven 3.6.3, which in turn used the \texttt{javac} compiler from OpenJDK 1.8.0\_442.
The tool also uses the ASM Java library to read the bytecode files (.class) in their binary format and uses the same Levenshtein distance to calculate the pairwise bytecode similarity between test cases.
A separate bytecode similarity CSV file is generated.
The textual and bytecode similarity files are to be used in Ledru-TCP.

To collect data for FAST-TCP, the tool saves the concatenated strings representing test cases and the hexadecimal form of the bytecodes in text files.
These text files are the input for FAST-TCP.

Additionally, the tool filters the bytecode instructions to include only instructions that push constant values onto the stack (like \texttt{iconst}), get field access instructions (like \texttt{getfield} or \texttt{putfield}), or invoke methods (like \texttt{invokevirtual} or \texttt{invokestatic}).
We used the ASM Java library to target these specific instructions and discard the rest.
As before, the tool produces separate filtered bytecode similarity files and text files that contain the hexadecimal form of the filtered bytecodes.

The tool obtains coverage, mutation, and error-revealing information about the test cases.
The tool ran each test case independently and used the generated JaCoCo reports to gather coverage information (statement and branch) for that test case.
The tool also runs PIT mutation analysis and parses the XML file generated by PIT to produce a mutation ``kill map''.
A {\it mutation ``kill map''} is a two-dimensional data structure that shows for each generated mutant which tests were able to kill that mutant.
The tool also runs the test suite against the buggy versions of Defects4J to record all the error-revealing tests.

\vspace{-0.1cm}

\subsubsection{Prioritising Test Cases}

We are now ready to perform our TCP strategies, which we described in Section~\ref{sec:TCP-approaches} and the coverage-based approaches described at the start of Section~\ref{sec:evaluation}.
We apply Ledru-TCP, described in Algorithm~\ref{algorithm:selection}, using three different similarity matrices: textual similarity, all bytecode similarity, and filtered bytecode similarity.
Also, we apply FAST-TCP, described in Algorithm~\ref{algorithm:FAST}, with three different input files, using the text of the test cases, the hexadecimal form of all bytecode, and the hexadecimal form of filtered bytecode.

\subsubsection{TCP Evaluation}

As described in Section~\ref{sec:eval-metrics}, we evaluate the TCP strategies through mutation score, real-fault detection, and runtime.
The tool uses the newly ordered test suite and the mutation information to calculate the APFD using Equation~\ref{eq:APFD} for each TCP strategy.

With every test case added to the list of ordered tests, the tool checks that test case against the list of error-revealing tests.
The tool records the index of the first test case to detect the real fault for each TCP strategy.
The tool records the time taken to generate the similarity matrix, build the required signatures, and the prioritisation time.

To validate our results, we employed the non-parametric ``Mann-Whitney U'' statistical test, as the data are independent and we cannot assume a normal distribution. 
Additionally, we used the Vargha-Delaney ($A_{12}$) effect size test~\cite{Vargha2000}, a non-parametric measure that compares two groups by estimating the probability that a randomly selected value from one group is larger than a randomly selected value from the other group.

\subsection{Threats to Validity}
\label{sec:threats-to-validity}

In this section, we address validity threats that can affect our results.
%

\textbf{Construct validity:} The selected string distance to measure test case similarity could be wrong.
We mitigated this by using a widely known distance metric (Levenshtein), which is suitable given the nature of the data as reported by other works~\cite{Elgendy2024, Elgendy2025, Ledru2012}.
Another threat would be that the implementation of Ledru-TCP and FAST-TCP can be invalid.
We mitigated this by running the implemented Ledru-TCP against the case studies provided by Ledru~\etal~\cite{Ledru2012}, getting the same results, and by using the replication package for FAST-TCP provided by Miranda~\etal~\cite{FAST2024}.
Miranda~\etal~\cite{Miranda2018} used the text of the entire test classes as records for the input, while we used the individual test cases as records.
We had an email correspondence with one of the authors to verify that we used the latest version of FAST, and that using individual test cases is valid.

\textbf{Internal validity:} The results can be affected by the nature of the used tests in our test suite, as the test suites we used are comprised of developer-written tests and automatically generated tests from Randoop and EvoSuite.
To mitigate this risk, we ran our experiments on different test suites, where we used developer-written tests only, Randoop tests only, and EvoSuite tests only.
We found negligible differences and our findings are not affected.
These results are reported in our replication package~\cite{Replication2025}.

\textbf{External validity:} The findings of our study may not generalise to different subjects.
We mitigated this risk by using seven different Java projects from Defects4J, a publicly available and widely used framework, and ran our experiments on \numversions versions of the projects and \uniqueclasses different classes within the projects.
The detailed results of each project are reported in our replication package.
Another threat is whether our results give a valid summary of performance.
To mitigate this risk, we go beyond single-dimensional summaries of performance (e.g., median) to include measures of variation (e.g., standard deviation), and confidence (e.g., using statistical testing and effect size).

\textbf{Reliability:} 
To address reproducibility concerns, we: (1) detailed all TCP approaches and subject characteristics, (2) thoroughly documented our methodology (including data collection and bytecode filtering criteria), and (3) provided a complete replication package on GitHub~\cite{Replication2025} with subjects, resources, implementation details, and an executable README.
%


\section{Results}
\label{sec:results}

\subsection{RQ1: Bytecode vs. Textual Diversity}

\begin{table}[t]
    \centering
    \begin{threeparttable}
        \caption{APFDs for all static TCP approaches across all projects. }
        \begin{tabular}{l@{~}r@{~~}r@{~~}r@{~~}r}
            \hline
            \textbf{Approach} & \textbf{ Median} & \textbf{SD} & \textbf{p-value} & \textbf{$A_{12}$} \\
            \hline

            \textbf{Ledru-Text}     & \allmedianLedrutextAPFD & \allsdLedrutextAPFD
                        & - & -
                        \\

\textbf{Ledru-Bytecode}     & \allmedianLedrubytecodeAPFD & \allsdLedrubytecodeAPFD
                        & 0.453 & 0.53*
                        \\

\textbf{Ledru-Bytecode-Filter}     & \textbf{\allmedianLedrubytecodefilterAPFD} & \allsdLedrubytecodefilterAPFD
                        & \textbf{0.040} & 0.59*
                        \\                        
\hline      
\textbf{FAST-Text}     & \allmedianFASTtextAPFD & \allsdFASTtextAPFD
                        & - & -
                        \\

\textbf{FAST-Bytecode}     & \textbf{\allmedianFASTbytecodeAPFD} & \allsdFASTbytecodeAPFD
                            & \textbf{0.000} & 0.67**
                        \\

\textbf{FAST-Bytecode-Filter}     & \allmedianFASTbytecodefilterAPFD & \allsdFASTbytecodefilterAPFD
                            & \textbf{0.013} & 0.60*
                        \\
\hline

\hline

        \end{tabular}
        \label{tab:APFD-static}
        \begin{tablenotes}
            \footnotesize
            \item SD is the standard deviation and $A_{12}$ is the effect size. The p-values and the effect sizes are between textual diversity and bytecode diversity for each TCP technique.
                A small effect size is marked with (*), while a medium effect size is marked with (**).
        \end{tablenotes}
    \end{threeparttable}
    \vspace{-0.2cm}
\end{table}

Tables~\ref{tab:APFD-static} and ~\ref{tab:realfault-static} refer to the APFDs and real-fault detection for the static diversity-based TCP approaches for all projects.
The first column is the TCP approach and diversity artefact used.
For instance, Ledru-Text applies Ledru-TCP using textual information, Ledru-Bytecode applies Ledru-TCP with all bytecode information, and FAST-Bytecode-Filter applies FAST-TCP employing the filtered bytecode information.
The next two columns have the median and standard deviation (SD).
The last two columns have the pairwise p-values and the Vargha-Delaney ($A_{12}$) effect sizes of the diversity-based TCP approaches.
The highest APFD median value of the TCP approaches and the lowest median number of the first test case to detect the real faults are highlighted in the table.
Also, we highlighted the p-values that are significantly different using $\alpha=0.05$ and marked different effect sizes with stars, where 1, 2, and 3 stars correspond to a small, medium or large effect size, respectively.
The p-values and effect sizes shown in the table are between the textual diversity and the corresponding bytecode diversity.

The median APFD of Ledru-Bytecode is 89.9, which is slightly higher than Ledru-Text.
The median real fault detection is 124, which means that on median, after 124 test cases run, the real fault can be detected.
This is slightly worse than the 110 tests of Ledru-Text.
However, the differences in the APFD and real fault detection values are not statistically significant and have small effect sizes over Ledru-Text.

\begin{table}[t]
    \centering
    
    \caption{Real-fault detection for all static TCP approaches across all projects. }
    \vspace{-0.2cm}
    \begin{tabular}{l@{~}r@{~~}r@{~~}r@{~~}r}
        \hline
        \textbf{Approach} & \textbf{ Median} & \textbf{SD} & \textbf{p-value} & \textbf{$A_{12}$} \\
        \hline

        \textbf{Ledru-Text}     & \allmedianLedrutextRealfault & \allsdLedrutextRealfault
                        & - & -
                        \\

\textbf{Ledru-Bytecode}     & \allmedianLedrubytecodeRealfault & \allsdLedrubytecodeRealfault
                            & 0.744 & 0.49*
                        \\

\textbf{Ledru-Bytecode-Filter}     & \textbf{\allmedianLedrubytecodefilterRealfault} & \allsdLedrubytecodefilterRealfault
                            & 0.211 & 0.45*
                        \\
\hline      
\textbf{FAST-Text}     & \allmedianFASTtextRealfault & \allsdFASTtextRealfault
                        & - & -
                        \\

\textbf{FAST-Bytecode}     & \textbf{\allmedianFASTbytecodeRealfault} & \allsdFASTbytecodeRealfault
                        & \textbf{0.007} & 0.61*
                        \\

\textbf{FAST-Bytecode-Filter}     & \allmedianFASTbytecodefilterRealfault & \allsdFASTbytecodefilterRealfault
                        & 0.297 & 0.46*
                        \\
\hline

\hline

    \end{tabular}
    \label{tab:realfault-static}
    \vspace{-0.3cm}
\end{table}

FAST-Bytecode has a median APFD of 93.2 and a median real fault detection of 32 (\ie, only 32 test cases to run to find the real fault).
The standard deviation is also the lowest, showing more consistency.
Both values are statistically significant at the 99\% significance level. 
APFD shows a medium effect size (0.67, which means that 67\% of the cases favored FAST-Bytecode over FAST-Text), while the real fault detection demonstrates a small effect size.

\begin{figure*}[htbp]
    \centering
    \begin{subfigure}[b]{0.33\textwidth}
        \centering
        \includegraphics[width=\textwidth]{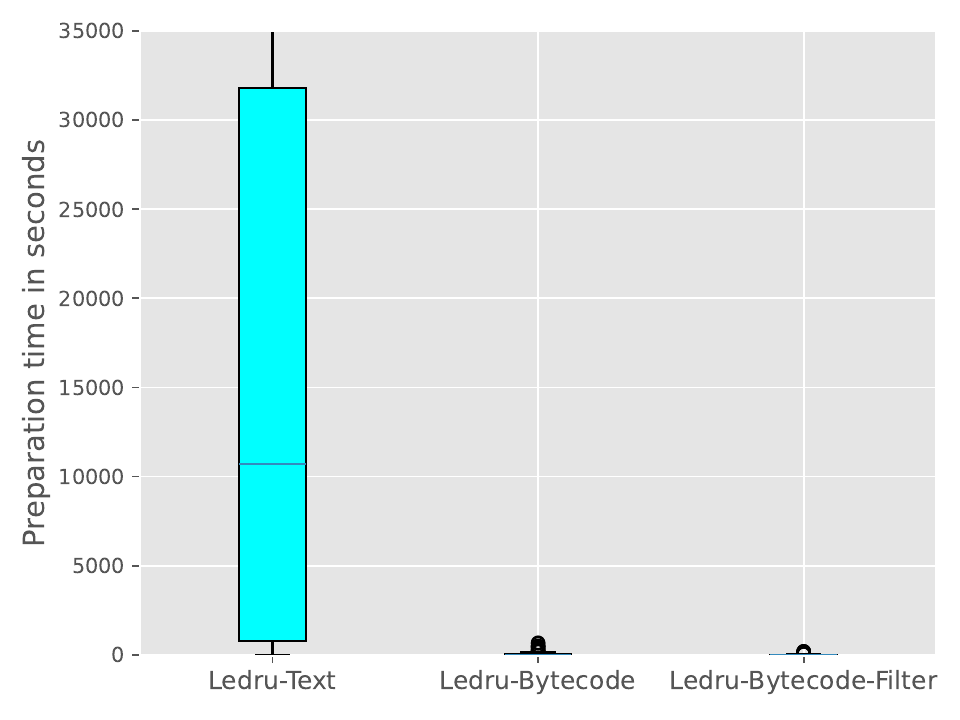} 
        \vspace{-2em}
        \caption{Ledru-TCP}
        \label{fig:prep-time1}
    \end{subfigure}
    \begin{subfigure}[b]{0.33\textwidth}
        \centering
        \includegraphics[width=\textwidth]{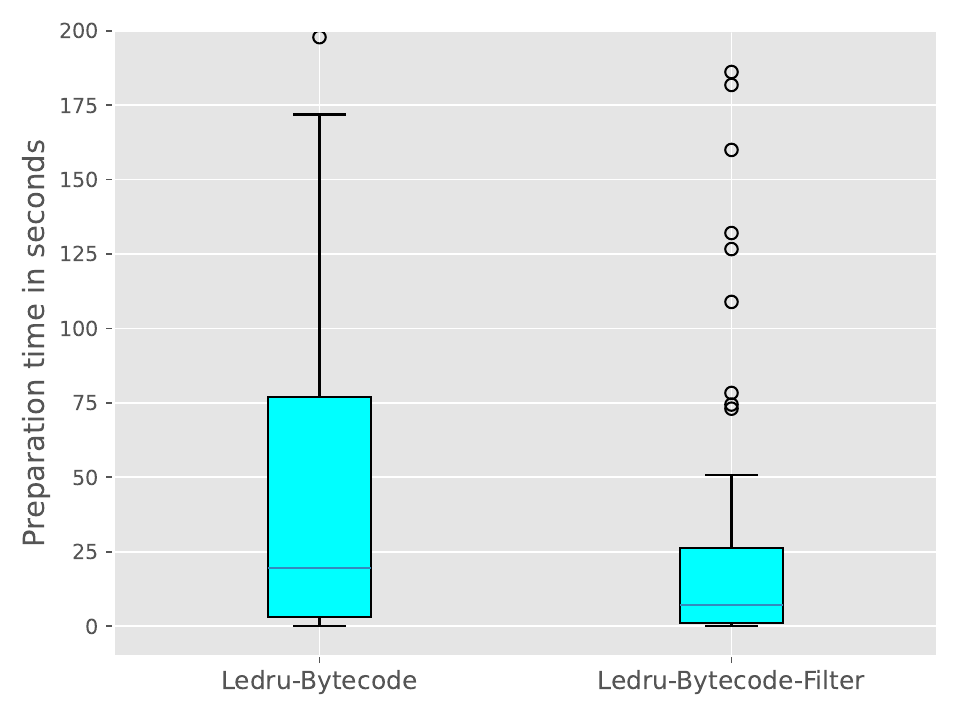} 
        \vspace{-2em}
        \caption{Ledru-TCP using bytecode}
        \label{fig:prep-time2}
    \end{subfigure}
    \begin{subfigure}[b]{0.33\textwidth}
        \centering
        \includegraphics[width=\textwidth]{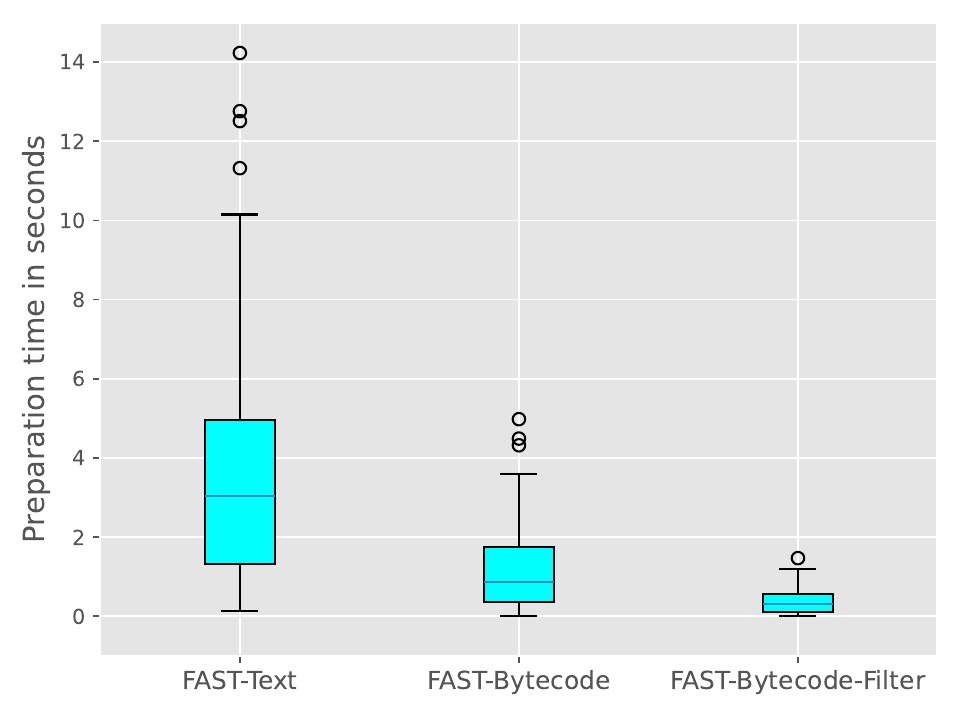} 
        \vspace{-2em}
        \caption{FAST-TCP}
        \label{fig:prep-time3}
    \end{subfigure}
    \vspace{-2em}
    \caption{The preparation times in seconds for all static TCP approaches}
    \label{fig:prep-time}
\end{figure*}

In terms of runtime, Figure~\ref{fig:prep-time} shows the preparation times for all static TCP approaches.
Table~\ref{tab:runtime} shows the runtime for all TCP approaches in each project in our study.
For Ledru-TCP and FAST-TCP, the diversity artefact has the biggest impact on the preparation time, while the prioritisation time is usually the same per TCP approach.
Thus, we reported the preparation times of each TCP approach and the average prioritisation time for Ledru-TCP and FAST-TCP.
Using bytecode rather than the text of test cases hugely improves the efficiency of the Ledru-TCP and FAST-TCP.

Using Ledru-TCP, the preparation times using all bytecode information were 124 to 814 times faster than using textual test cases, where the bigger the test suite size and the larger the test case, the more improvements can be observed.
%
Figure~\ref{fig:prep-time1} shows a substantial difference between using text and bytecode in Ledru-TCP. 
Similarly, Figure~\ref{fig:prep-time3} highlights a significant reduction in preparation time when using bytecode instead of text in FAST-TCP. 
The smallest project in our study is the $\texttt{Time}$ project, where the test suite size is 740 test cases.
The preparation time in Ledru-TCP for building the similarity matrix dropped from 633.3 seconds (almost 10 minutes) using text to 5.1 seconds using all bytecode information.
In FAST-TCP, the preparation time for building the signatures dropped from 2.6 seconds to 0.6 seconds.
The $\texttt{Math}$ project has the largest number of different versions and the largest number of test cases (29,610), but the test cases are small.
In Ledru-TCP, the preparation times dropped from 347037.9 seconds (almost 96 hours) to 757.7 seconds (almost 12 minutes), while in FAST-TCP, the preparation times dropped from 88.9 seconds to 33.2 seconds.
In FAST-TCP, the preparation times using all bytecode information were 2.5 to 6 times faster than using the text of test cases.

\begin{table*}[htbp]
    \centering
    \begin{threeparttable}
        \caption{Runtime for all TCP approaches across all projects. }
        
        \begin{tabular}{l@{~}|r@{~~}r@{~~}r@{~~}r|@{~~}r@{~~}r@{~~}r@{~~}r@{~~}|r@{~~}r}
            \hline
            \textbf{Project} & \makecell{\textbf{Ledru-}\\ \textbf{Text}} & \makecell{\textbf{Ledru-} \\ \textbf{Bytecode}} & \makecell{\textbf{Ledru-} \\ \textbf{Bytecode-} \\ \textbf{Filter}} 
            & \makecell{\textbf{Ledru Avg} \\ \textbf{P-Time}} & \makecell{\textbf{FAST-} \\ \textbf{Text}} & \makecell{\textbf{FAST-} \\ \textbf{Bytecode}} & \makecell{\textbf{FAST-} \\ \textbf{Bytecode-} \\ \textbf{Filter}} & \makecell{\textbf{FAST Avg} \\ \textbf{P-Time}}
            & \makecell{\textbf{Cov-Tot}} & \makecell{\textbf{Cov-Add}}\\
            \hline

            \textbf{Cli}    & \clitextPreptime 
                & \clibytecodePreptime 
                & \clibytecodefilterPreptime 
                & \cliAvgLedruPtime 
                & \clifastPreptime 
                & \clifastbytecodePreptime 
                & \clifastbytecodefilterPreptime 
                & \cliAvgFASTPtime 
                & \cliCovTotTotaltime
                & \cliCovAddTotaltime \\
\hline
\textbf{Compress}   & \compresstextPreptime 
                    & \compressbytecodePreptime 
                    & \compressbytecodefilterPreptime 
                    & \compressAvgLedruPtime 
                    & \compressfastPreptime 
                    & \compressfastbytecodePreptime 
                    & \compressfastbytecodefilterPreptime 
                    & \compressAvgFASTPtime 
                    & \compressCovTotTotaltime
                    & \compressCovAddTotaltime \\
\hline
\textbf{Csv}    & \csvtextPreptime 
                & \csvbytecodePreptime 
                & \csvbytecodefilterPreptime 
                & \csvAvgLedruPtime 
                & \csvfastPreptime 
                & \csvfastbytecodePreptime 
                & \csvfastbytecodefilterPreptime 
                & \csvAvgFASTPtime 
                & \csvCovTotTotaltime
                & \csvCovAddTotaltime \\
\hline
\textbf{Jsoup}  & \jsouptextPreptime 
                & \jsoupbytecodePreptime 
                & \jsoupbytecodefilterPreptime 
                & \jsoupAvgLedruPtime 
                & \jsoupfastPreptime 
                & \jsoupfastbytecodePreptime 
                & \jsoupfastbytecodefilterPreptime 
                & \jsoupAvgFASTPtime 
                & \jsoupCovTotTotaltime
                & \jsoupCovAddTotaltime \\
\hline
\textbf{Lang}   & \langtextPreptime 
                & \langbytecodePreptime 
                & \langbytecodefilterPreptime 
                & \langAvgLedruPtime 
                & \langfastPreptime 
                & \langfastbytecodePreptime 
                & \langfastbytecodefilterPreptime 
                & \langAvgFASTPtime 
                & \langCovTotTotaltime
                & \langCovAddTotaltime \\
\hline
\textbf{Math}   & \mathtextPreptime 
                & \mathbytecodePreptime 
                & \mathbytecodefilterPreptime 
                & \mathAvgLedruPtime 
                & \mathfastPreptime 
                & \mathfastbytecodePreptime 
                & \mathfastbytecodefilterPreptime 
                & \mathAvgFASTPtime 
                & \mathCovTotTotaltime
                & \mathCovAddTotaltime \\
\hline
\textbf{Time}   & \timetextPreptime 
                & \timebytecodePreptime 
                & \timebytecodefilterPreptime 
                & \timeAvgLedruPtime 
                & \timefastPreptime 
                & \timefastbytecodePreptime 
                & \timefastbytecodefilterPreptime 
                & \timeAvgFASTPtime 
                & \timeCovTotTotaltime
                & \timeCovAddTotaltime \\
\hline

        \end{tabular}
        \label{tab:runtime}
        \begin{tablenotes}
            \footnotesize
            \item Avg P-Time is the average prioritisation time in seconds for the TCP approach. Cov-Tot is coverage-based greedy total, while Cov-Add is coverage-based greedy additional.
        \end{tablenotes}
    \end{threeparttable}
    \vspace{-0.2cm}
\end{table*}

\begin{framed}
   \noindent \textbf{Conclusion for RQ1:} The experiments show that bytecode diversity has higher APFDs than textual diversity, as it increased by 2.3\% in Ledru-TCP and increased by 7.8\% in FAST-TCP.
   The real fault detection in Ledru-Text was slightly better than Ledru-Bytecode, but FAST-Bytecode was 3.9 times better than FAST-Text.
   The most significant improvement was in efficiency, as using bytecode information is 124 to 814 times faster than using textual information in Ledru-TCP, while in FAST-TCP, bytecode is 2.5 to 6 times faster than text.
   This dramatic difference in runtime means that bytecode-based approaches can complete the entire prioritisation process while Ledru-Text is still calculating similarity values. Although textual diversity shows a marginal advantage in real fault detection for Ledru-TCP, the enormous runtime savings of bytecode diversity make it a more practical and scalable choice for large-scale regression testing.
\end{framed}

\subsection{RQ2: Filtering Bytecode}

Filtering bytecode instructions has a positive effect in Ledru-TCP, as the median APFD increased from 87.6 in text to 92.1 using filtered bytecode information.
This is also an increase of 2.2\% over using all bytecode information.
Furthermore, it is statistically significant at the 95\% significance level and has a small effect size in favour of filtered bytecode over using text.
Additionally, the median fault detection of Ledru-Bytecode-Filter is 78, which is 1.4 times better than Ledru-Text and 1.6 times better than Ledru-Bytecode.

In FAST-TCP, using filtered bytecode improved APFD by 5.1\% compared with text, but it was 2.7\% lower than all bytecode information.
%
FAST-Bytecode-Filter demonstrates a statistically significant improvement over FAST-Text at the 95\% significance level, with a small effect size.
The median real fault detection is 63, which is almost twice as good as FAST-Text.
However, the difference is not statistically significant and has a small effect size.

Once more, the biggest improvement is the efficiency of filtering bytecode information.
Ledru-Bytecode-Filter is 455 to 2798 times faster than Ledru-Text and 2.5 to 4.1 times faster than Ledru-Bytecode.
Figures~\ref{fig:prep-time2}-~\ref{fig:prep-time3} visualise these time reductions for both full and filtered bytecode approaches.
%
FAST-Bytecode-Filter is 4 to 17.9 times faster than FAST-Text and 1.4 to 4 times faster than FAST-Bytecode.
%
%
The preparation time in the $\texttt{Math}$ project dropped to only 261.4 seconds (4 minutes) in Ledru-TCP, while it dropped to only 9.2 seconds in FAST-TCP.

\begin{framed}
   \noindent \textbf{Conclusion for RQ2:} Filtering bytecode information increased the APFD in Ledru-TCP by 4.5\%, and it was 1.4 times better in real fault detection than using the text of test cases.
   In FAST-TCP, the APFD increased by 5.1\% and the real fault detection was twice as good as using text of test cases.
   Once more, the biggest improvement was in efficiency, where in Ledru it was 455 to 2798 times faster than Ledru-Text and 2.5 to 4.1 times faster than Ledru-Bytecode.
   In FAST-TCP, filtering bytecode was 4 to 17.9 times faster than text and 1.4 to 4 times faster than all bytecode information.
\end{framed}

\subsection{RQ3: Bytecode TCP vs. Dynamic TCP}

We compared the best-performing static TCP approach (FAST-Bytecode) with the dynamic coverage-based TCP approaches.
Tables~\ref{tab:APFD-dynamic} and~\ref{tab:Realfaults-dynamic} show the APFDs and real fault detection results.
%
%
The median APFD of greedy additional is 96.7, which is slightly more than the greedy total (96), but the difference is not statistically significant.
FAST-Bytecode has lower APFD by 2.8\% to 3.5\% than greedy total and greedy additional, respectively.
Also, the standard deviation for the two coverage-based approaches is lower than bytecode diversity showing better consistency.
Furthermore, the results are statistically significant at the 99\% significance level, and there is a medium effect size for coverage-based TCP approaches.
%

\begin{table}[t]
    \centering

        \caption{APFDs for FAST-bytecode TCP and coverage-based TCP across all projects. }
        \vspace{-0.2cm}
        \begin{tabular}{l@{~}r@{~~}r@{~~}r@{~~}r}
            \hline
            \textbf{Approach} & \textbf{ Median} & \textbf{SD} & \textbf{p-value} & \textbf{$A_{12}$} \\
            \hline

\textbf{FAST-Bytecode}      & \allmedianFASTbytecodeAPFD & \allsdFASTbytecodeAPFD
                            & - & -
                        \\

\textbf{Cov-Total}      & \allmediancovTot & \allsdcovTot
                        & \textbf{0.003} & 0.62**
                        \\

\textbf{Cov-Additional}     & \textbf{\allmediancovAdd} & \allsdcovAdd
                        & \textbf{0.001} & 0.64**
                        \\

\hline

        \end{tabular}
        \label{tab:APFD-dynamic}
        \vspace{-0.2cm}
\end{table}

\begin{table}[t]
    \centering
        \caption{APFDs for FAST-bytecode TCP and coverage-based TCP across all projects. }
        \vspace{-0.2cm}
        \begin{tabular}{l@{~}r@{~~}r@{~~}r@{~~}r}
            \hline
            \textbf{Approach} & \textbf{ Median} & \textbf{SD} & \textbf{p-value} & \textbf{$A_{12}$} \\
            \hline

\textbf{FAST-Bytecode}      & \allmedianFASTbytecodeRealfault & \allsdFASTbytecodeRealfault
                            & - & -
                        \\

\textbf{Cov-Total}      & \allmediancovTotRealfault & \allsdcovTotRealfault
                        & \textbf{0.011} & 0.61*
                        \\

\textbf{Cov-Additional}     & \textbf{\allmediancovAddRealfault} & \allsdcovAddRealfault
                        & \textbf{0.000} & 0.65**
                        \\

\hline

        \end{tabular}
        \label{tab:Realfaults-dynamic}
        \vspace{-0.2cm}
\end{table}

In terms of real fault detection, the median numbers of the first test case to detect the real fault of greedy additional and greedy total are 9 and 10.5, respectively.
The results of coverage-based TCP approaches are statistically significant at the 95\% significance level over bytecode diversity, and there is a medium effect size in favour of bytecode diversity.

Finally, the overall time of FAST-Bytecode, which includes the preparation and prioritisation times, is better than coverage-based TCP.
%
%
As explained earlier in Section~\ref{sec:eval-metrics}, we included only the prioritisation times for greedy total and greedy additional, shown in Table~\ref{tab:runtime}.
Nonetheless, even using prioritisation times only for coverage-based TCP, FAST-Bytecode is still up to 6.5 times faster than greedy total and up to 8.3 times faster than greedy additional.
In the $\texttt{Jsoup}$ project, the overall time using greedy total is 446.1 seconds (7.4 minutes) and using greedy additional was 568.7 seconds (9.5 minutes), while using FAST-Bytecode the overall time is 68.6 seconds (1.1 minutes).

\begin{framed}
   \noindent \textbf{Conclusion for RQ3:} Coverage-based TCP outperform bytecode diversity TCP approaches, and there is little difference between greedy total and greedy additional.
   The results are statistically significant at the 95\% significance level between bytecode and coverage-based approaches.
   However, the total runtime using FAST-Bytecode is up to 6.5 times faster than greedy total and up to 8.3 times faster than greedy additional.
   For larger test suites, the difference in runtime becomes larger.
\end{framed}

\section{Related Work}
\label{sec:related-work}

There is a plethora of work in TCP in general, and using diversity-based TCP in particular.
This shows how TCP is a very active topic in the software testing literature.
Yoo~\etal~\cite{Yoo2010B} presented a comprehensive survey of techniques and strategies used in regression testing, where TCP is one of the techniques described to deal with regression testing.
They categorised the TCP techniques as coverage-based (\ie, use structural coverage as metric), interaction testing (\ie, use multiple combinations of different components), distribution-based (\ie, use the distributions of the profiles of test cases),
human-based (\ie, use case-based reasoning), history-based (\ie, use past bug fixes records), requirement-based (\ie, use software requirements), model-based (\ie, use system models), and other approaches (e.g. use of mutation scores, or use of recorded sessions for web applications).
Lou~\etal~\cite{Lou2019} presented 191 papers on TCP from 1997 to 2016, and investigated six aspects, which are the algorithms, criteria, measurements, constraints, empirical studies, and scenarios.
Elgendy~\etal~\cite{Elgendy2025} provided a comprehensive survey of diversity-based testing in software testing and covered diversity-based techniques in TCP.
They identified text, test steps (\ie, individual actions that need to be followed to execute a test case), program executions, and clustering as common diversity artefacts in TCP but highlighted the need to explore new diversity artefacts. 
Our work directly addresses this gap by introducing bytecode as a novel diversity artefact for TCP, offering a more concise and efficient representation of test cases compared to traditional textual diversity.

Several researchers have developed TCP strategies to enhance early fault detection and test suite efficiency. 
Wu~\etal~\cite{Wu2012} improved similarity-based prioritisation by incorporating program element execution times, while Khojah~\etal~\cite{Khojah2023} compared lexical and semantic diversity, concluding that semantic diversity provides better requirement coverage. 
Altiero~\etal~\cite{Altiero2024} prioritised tests based on their coverage of modified code parts, and Ledru~\etal~\cite{Ledru2012} introduced a greedy algorithm to maximise test diversity using textual representations.
While these studies demonstrate the effectiveness of diversity-based TCP, they rely on textual diversity. 
In contrast, our work uses textual diversity as a baseline for comparison with bytecode diversity.

For manual black-box system testing, Hemmati~\etal~\cite{Hemmati2015} adapted test prioritisation techniques, and empirical studies have evaluated the effectiveness of similarity-based TCP.
Huang~\etal~\cite{Huang2017} compared local and global TCP techniques, finding global approaches superior in coverage and fault detection.
Haghighatkhah~\etal~\cite{Haghighatkhah2018A} identified the most effective similarity-based TCP implementation, testing various distance measures.
Noor and Hemmati~\cite{Noor2015} proposed prioritising test cases by their similarity to previously failing test cases, using metrics like Basic Counting, Hamming distance, and Levenshtein distance, suggesting that tests similar to past failures are more likely to detect new faults.
In addressing scalability and performance for large test suites, Miranda~\etal~\cite{Miranda2018} introduced the FAST family of techniques, which leverage big data methods like Shingling, Minhashing, and Locality Sensitive Hashing (LSH) for efficient similarity detection in both white-box and black-box testing.
We implemented our bytecode diversity TCP and compared it against Ledru~\etal~\cite{Ledru2012} and FAST~\cite{Miranda2018}.

\section{Conclusions and Future Work}
\label{sec:conclusions-and-future-work}

This study introduces and evaluates, for the first time, the use of bytecode (a more concise representation of test cases) as the basis for Test Case Prioritisation (TCP). 
%
%
The results show that bytecode diversity improves fault detection rates and significantly reduces processing times compared to traditional textual diversity-based methods.
Also, the results show that filtering bytecode information greatly improves efficiency while achieving better fault detection than textual information and comparable effectiveness to using all bytecode information.
Although coverage-based techniques yield better fault detection, they require extensive program instrumentation, making them less practical in certain contexts.
Bytecode diversity thus provides a suitable alternative, offering a balance between effectiveness and operational simplicity.
Our findings suggest that bytecode-based TCP can be particularly advantageous in resource-constrained environments or where coverage data is unavailable.

Future research can explore different ways to filter bytecode information and gain insights into which instructions can be the most beneficial in TCP.
One potential enhancement to our approach and an avenue for future work is using normalised bytecode for similarity analysis. 
Raw bytecode can vary due to compiler optimisations and structural differences that do not affect program behaviour. 
Techniques such as bytecode canonicalisation, variable renaming, and control flow normalisation (e.g., SootDiff~\etal~\cite{Dann2019}, Schott~\etal~\cite{Schott2024}) could help filter out these variations, leading to more precise similarity calculations. 
We could also conduct experiments on datasets having multi-fault programs and use the cost-cognizant version of the APFD metric to investigate different fault severity.
Additionally, a deeper investigation into diversity-based approaches, particularly focusing on the effectiveness of bytecode diversity in addressing stubborn mutants (those hardest to detect/kill) could provide valuable insights.
Such work would further validate the robustness of bytecode diversity and its potential to improve the fault detection capabilities of regression testing.

\balance
\bibliography{bibliography} 


\begin{thebibliography}{35}


\ifx \showCODEN    \undefined \def \showCODEN     #1{\unskip}     \fi
\ifx \showDOI      \undefined \def \showDOI       #1{#1}\fi
\ifx \showISBNx    \undefined \def \showISBNx     #1{\unskip}     \fi
\ifx \showISBNxiii \undefined \def \showISBNxiii  #1{\unskip}     \fi
\ifx \showISSN     \undefined \def \showISSN      #1{\unskip}     \fi
\ifx \showLCCN     \undefined \def \showLCCN      #1{\unskip}     \fi
\ifx \shownote     \undefined \def \shownote      #1{#1}          \fi
\ifx \showarticletitle \undefined \def \showarticletitle #1{#1}   \fi
\ifx \showURL      \undefined \def \showURL       {\relax}        \fi
\providecommand\bibfield[2]{#2}
\providecommand\bibinfo[2]{#2}
\providecommand\natexlab[1]{#1}
\providecommand\showeprint[2][]{arXiv:#2}

\bibitem[Altiero et~al\mbox{.}(2024)]%
        {Altiero2024}
\bibfield{author}{\bibinfo{person}{F. Altiero}, \bibinfo{person}{A. Corazza},
  \bibinfo{person}{S. Di~Martino}, \bibinfo{person}{A. Peron}, {and}
  \bibinfo{person}{L. Libero Lucio~Starace}.} \bibinfo{year}{2024}\natexlab{}.
\newblock \showarticletitle{Regression test prioritization leveraging source
  code similarity with tree kernels}.
\newblock \bibinfo{journal}{\emph{Journal of Software: Evolution and Process}}
  (\bibinfo{year}{2024}), \bibinfo{pages}{e2653}.
\newblock
\urldef\tempurl%
\url{https://doi.org/10.1002/smr.2653}
\showDOI{\tempurl}


\bibitem[Arafeen and Do(2013)]%
        {Arafeen2013}
\bibfield{author}{\bibinfo{person}{Md~J. Arafeen} {and} \bibinfo{person}{H.
  Do}.} \bibinfo{year}{2013}\natexlab{}.
\newblock \showarticletitle{Test case prioritization using requirements-based
  clustering}. In \bibinfo{booktitle}{\emph{Proceedings of the International
  Conference on Software Testing, Validation and Verification (ICST)}}.
  \bibinfo{pages}{312--321}.
\newblock
\urldef\tempurl%
\url{https://doi.org/10.1109/ICST.2013.12}
\showDOI{\tempurl}


\bibitem[Coles et~al\mbox{.}(2016)]%
        {Coles2016}
\bibfield{author}{\bibinfo{person}{H. Coles}, \bibinfo{person}{T. Laurent},
  \bibinfo{person}{C. Henard}, \bibinfo{person}{M. Papadakis}, {and}
  \bibinfo{person}{A. Ventresque}.} \bibinfo{year}{2016}\natexlab{}.
\newblock \showarticletitle{PIT: A practical mutation testing tool for java}.
  In \bibinfo{booktitle}{\emph{Proceedings of the International Symposium on
  Software Testing and Analysis (ISSTA)}}. \bibinfo{pages}{449--452}.
\newblock
\urldef\tempurl%
\url{https://doi.org/10.1145/2931037.2948707}
\showDOI{\tempurl}


\bibitem[Dann et~al\mbox{.}(2019)]%
        {Dann2019}
\bibfield{author}{\bibinfo{person}{A. Dann}, \bibinfo{person}{B. Hermann},
  {and} \bibinfo{person}{E. Bodden}.} \bibinfo{year}{2019}\natexlab{}.
\newblock \showarticletitle{SootDiff: bytecode comparison across different Java
  compilers}. In \bibinfo{booktitle}{\emph{Proceedings of the International
  Workshop on State Of the Art in Program Analysis}}. \bibinfo{pages}{14–19}.
\newblock
\urldef\tempurl%
\url{https://doi.org/10.1145/3315568.3329966}
\showDOI{\tempurl}


\bibitem[Elbaum et~al\mbox{.}(2001)]%
        {Elbaum2001}
\bibfield{author}{\bibinfo{person}{S. Elbaum}, \bibinfo{person}{A.
  Malishevsky}, {and} \bibinfo{person}{G. Rothermel}.}
  \bibinfo{year}{2001}\natexlab{}.
\newblock \showarticletitle{Incorporating varying test costs and fault
  severities into test case prioritization}. In
  \bibinfo{booktitle}{\emph{Proceedings of the International Conference on
  Software Engineering (ICSE)}}. \bibinfo{pages}{329--338}.
\newblock
\urldef\tempurl%
\url{https://doi.org/10.1109/ICSE.2001.919106}
\showDOI{\tempurl}


\bibitem[Elgendy et~al\mbox{.}(2024)]%
        {Elgendy2024}
\bibfield{author}{\bibinfo{person}{I. Elgendy}, \bibinfo{person}{R. Hierons},
  {and} \bibinfo{person}{P. McMinn}.} \bibinfo{year}{2024}\natexlab{}.
\newblock \showarticletitle{Evaluating string distance metrics for reducing
  automatically generated test suites}. In
  \bibinfo{booktitle}{\emph{Proceedings of the International Conference on
  Automation of Software Test (AST)}}. \bibinfo{pages}{171--181}.
\newblock
\urldef\tempurl%
\url{https://doi.org/10.1145/3644032.3644455}
\showDOI{\tempurl}


\bibitem[Elgendy et~al\mbox{.}(2025a)]%
        {Replication2025}
\bibfield{author}{\bibinfo{person}{I. Elgendy}, \bibinfo{person}{R. Hierons},
  {and} \bibinfo{person}{P. McMinn}.} \bibinfo{year}{2025}\natexlab{a}.
\newblock \bibinfo{title}{{Replication Package for Empirically Evaluating the
  Use of Bytecode for Diversity-Based Test Case Prioritisation}}.
\newblock
  \bibinfo{howpublished}{\url{https://github.com/islamelgendy/Replication-Package-Evaluating-Bytecode-Diversity}}.
\newblock
\newblock
\shownote{[Online; accessed 25-March-2025]}.


\bibitem[Elgendy et~al\mbox{.}(2025b)]%
        {Elgendy2025}
\bibfield{author}{\bibinfo{person}{I. Elgendy}, \bibinfo{person}{R. Hierons},
  {and} \bibinfo{person}{P. McMinn}.} \bibinfo{year}{2025}\natexlab{b}.
\newblock \showarticletitle{A Systematic Mapping Study of the Metrics, Uses and
  Subjects of Diversity-Based Testing Techniques}.
\newblock \bibinfo{journal}{\emph{Software Testing, Verification and
  Reliability}} \bibinfo{volume}{35}, \bibinfo{number}{2}
  (\bibinfo{year}{2025}), \bibinfo{pages}{e1914}.
\newblock
\urldef\tempurl%
\url{https://doi.org/10.1002/stvr.1914}
\showDOI{\tempurl}


\bibitem[Eric et~al\mbox{.}(2002)]%
        {Asm2025}
\bibfield{author}{\bibinfo{person}{B. Eric}, \bibinfo{person}{L. Romain}, {and}
  \bibinfo{person}{C. Thierry}.} \bibinfo{year}{2002}\natexlab{}.
\newblock \bibinfo{title}{ASM: A code manipulation tool for the Java virtual
  machine}.
\newblock
\newblock
\newblock
\shownote{Available: \url{https://asm.ow2.io}}.


\bibitem[Fraser and Arcuri(2011)]%
        {Fraser2011}
\bibfield{author}{\bibinfo{person}{G. Fraser} {and} \bibinfo{person}{A.
  Arcuri}.} \bibinfo{year}{2011}\natexlab{}.
\newblock \showarticletitle{Evosuite: automatic test suite generation for
  object-oriented software}. In \bibinfo{booktitle}{\emph{Proceedings of the
  SIGSOFT symposium and the European conference on Foundations of software
  engineering}}. \bibinfo{pages}{416--419}.
\newblock
\urldef\tempurl%
\url{https://doi.org/10.1145/2025113.2025179}
\showDOI{\tempurl}


\bibitem[Haghighatkhah et~al\mbox{.}(2018a)]%
        {Haghighatkhah2018A}
\bibfield{author}{\bibinfo{person}{A Haghighatkhah}, \bibinfo{person}{M.
  M{\"a}ntyl{\"a}}, \bibinfo{person}{M. Oivo}, {and} \bibinfo{person}{P.
  Kuvaja}.} \bibinfo{year}{2018}\natexlab{a}.
\newblock \showarticletitle{Test case prioritization using test similarities}.
  In \bibinfo{booktitle}{\emph{Proceedings of the International Conference on
  Product-Focused Software Process Improvement (PROFES)}}.
  \bibinfo{pages}{243--259}.
\newblock
\urldef\tempurl%
\url{https://doi.org/10.1007/978-3-030-03673-7\_18}
\showDOI{\tempurl}


\bibitem[Haghighatkhah et~al\mbox{.}(2018b)]%
        {Haghighatkhah2018B}
\bibfield{author}{\bibinfo{person}{A. Haghighatkhah}, \bibinfo{person}{M.
  M{\"a}ntyl{\"a}}, \bibinfo{person}{M. Oivo}, {and} \bibinfo{person}{P.
  Kuvaja}.} \bibinfo{year}{2018}\natexlab{b}.
\newblock \showarticletitle{Test prioritization in continuous integration
  environments}.
\newblock \bibinfo{journal}{\emph{Journal of Systems and Software}}
  \bibinfo{volume}{146} (\bibinfo{year}{2018}), \bibinfo{pages}{80--98}.
\newblock
\urldef\tempurl%
\url{https://doi.org/10.1016/j.jss.2018.08.061}
\showDOI{\tempurl}


\bibitem[Hemmati et~al\mbox{.}(2015)]%
        {Hemmati2015}
\bibfield{author}{\bibinfo{person}{H. Hemmati}, \bibinfo{person}{Z. Fang},
  {and} \bibinfo{person}{M.~V Mantyla}.} \bibinfo{year}{2015}\natexlab{}.
\newblock \showarticletitle{Prioritizing manual test cases in traditional and
  rapid release environments}. In \bibinfo{booktitle}{\emph{Proceedings of the
  International Conference on Software Testing, Verification and Validation
  (ICST)}}. \bibinfo{pages}{1--10}.
\newblock
\urldef\tempurl%
\url{https://doi.org/10.1109/ICST.2015.7102602}
\showDOI{\tempurl}


\bibitem[Henard et~al\mbox{.}(2016)]%
        {Henard2016}
\bibfield{author}{\bibinfo{person}{C. Henard}, \bibinfo{person}{M. Papadakis},
  \bibinfo{person}{M. Harman}, \bibinfo{person}{Y. Jia}, {and}
  \bibinfo{person}{Y. Le~Traon}.} \bibinfo{year}{2016}\natexlab{}.
\newblock \showarticletitle{Comparing white-box and black-box test
  prioritization}. In \bibinfo{booktitle}{\emph{Proceedings of the
  International Conference on Software Engineering}}.
  \bibinfo{pages}{523--534}.
\newblock
\urldef\tempurl%
\url{https://doi.org/10.1145/2884781.2884791}
\showDOI{\tempurl}


\bibitem[Huang et~al\mbox{.}(2017)]%
        {Huang2017}
\bibfield{author}{\bibinfo{person}{R. Huang}, \bibinfo{person}{Y. Zhou},
  \bibinfo{person}{W. Zong}, \bibinfo{person}{D. Towey}, {and}
  \bibinfo{person}{J. Chen}.} \bibinfo{year}{2017}\natexlab{}.
\newblock \showarticletitle{An empirical comparison of similarity measures for
  abstract test case prioritization}. In \bibinfo{booktitle}{\emph{Proceedings
  of the Annual Computer Software and Applications Conference (COMPSAC)}},
  Vol.~\bibinfo{volume}{1}. \bibinfo{pages}{3--12}.
\newblock


\bibitem[JaCoCo(2025)]%
        {JaCoCo2024}
\bibfield{author}{\bibinfo{person}{JaCoCo}.} \bibinfo{year}{2025}\natexlab{}.
\newblock \bibinfo{title}{JaCoCo Implementation Design}.
\newblock
  \bibinfo{howpublished}{\url{http://www.jacoco.org/jacoco/trunk/doc/implementation.html}}.
\newblock
\newblock
\shownote{[Last accessed: 25-March-2025]}.


\bibitem[Just et~al\mbox{.}(2014)]%
        {Just2014A}
\bibfield{author}{\bibinfo{person}{R. Just}, \bibinfo{person}{D. Jalali}, {and}
  \bibinfo{person}{M.~D Ernst}.} \bibinfo{year}{2014}\natexlab{}.
\newblock \showarticletitle{Defects4J: A database of existing faults to enable
  controlled testing studies for Java programs}. In
  \bibinfo{booktitle}{\emph{Proceedings of the International Symposium on
  Software Testing and Analysis (ISSTA)}}. \bibinfo{pages}{437--440}.
\newblock
\urldef\tempurl%
\url{https://doi.org/10.1145/2610384.2628055}
\showDOI{\tempurl}


\bibitem[Khatibsyarbini et~al\mbox{.}(2018)]%
        {Khatibsyarbini2018}
\bibfield{author}{\bibinfo{person}{M. Khatibsyarbini}, \bibinfo{person}{M~A.
  Isa}, \bibinfo{person}{D.~NA Jawawi}, {and} \bibinfo{person}{R. Tumeng}.}
  \bibinfo{year}{2018}\natexlab{}.
\newblock \showarticletitle{Test case prioritization approaches in regression
  testing: A systematic literature review}.
\newblock \bibinfo{journal}{\emph{Information and Software Technology}}
  \bibinfo{volume}{93} (\bibinfo{year}{2018}), \bibinfo{pages}{74--93}.
\newblock
\urldef\tempurl%
\url{https://doi.org/10.1016/j.infsof.2017.08.014}
\showDOI{\tempurl}


\bibitem[Khojah et~al\mbox{.}(2023)]%
        {Khojah2023}
\bibfield{author}{\bibinfo{person}{R. Khojah}, \bibinfo{person}{C.~H. Chao},
  {and} \bibinfo{person}{d~F~G Oliveira~Neto}.}
  \bibinfo{year}{2023}\natexlab{}.
\newblock \showarticletitle{Evaluating the trade-offs of text-based diversity
  in test prioritisation}. In \bibinfo{booktitle}{\emph{Proceedings of the
  International Conference on Automation of Software Test (AST)}}.
  \bibinfo{pages}{168--178}.
\newblock
\urldef\tempurl%
\url{https://doi.org/10.1109/AST58925.2023.00021}
\showDOI{\tempurl}


\bibitem[Kim and Porter(2002)]%
        {Kim2002}
\bibfield{author}{\bibinfo{person}{J-M Kim} {and} \bibinfo{person}{A. Porter}.}
  \bibinfo{year}{2002}\natexlab{}.
\newblock \showarticletitle{A history-based test prioritization technique for
  regression testing in resource constrained environments}. In
  \bibinfo{booktitle}{\emph{Proceedings of the International Conference on
  Software Engineering (ICSE)}}. \bibinfo{pages}{119--129}.
\newblock
\urldef\tempurl%
\url{https://doi.org/10.1145/581339.581357}
\showDOI{\tempurl}


\bibitem[Ledru et~al\mbox{.}(2012)]%
        {Ledru2012}
\bibfield{author}{\bibinfo{person}{Y. Ledru}, \bibinfo{person}{A. Petrenko},
  \bibinfo{person}{S. Boroday}, {and} \bibinfo{person}{N. Mandran}.}
  \bibinfo{year}{2012}\natexlab{}.
\newblock \showarticletitle{Prioritizing test cases with string distances}.
\newblock \bibinfo{journal}{\emph{Automated Software Engineering}}
  \bibinfo{volume}{19}, \bibinfo{number}{1} (\bibinfo{year}{2012}),
  \bibinfo{pages}{65--95}.
\newblock
\urldef\tempurl%
\url{https://doi.org/10.1007/s10515-011-0093-0}
\showDOI{\tempurl}


\bibitem[Levenshtein(1966)]%
        {Levenshtein1966}
\bibfield{author}{\bibinfo{person}{V.~I. Levenshtein}.}
  \bibinfo{year}{1966}\natexlab{}.
\newblock \showarticletitle{Binary codes capable of correcting deletions,
  insertions, and reversals}.
\newblock  \bibinfo{volume}{10}, \bibinfo{number}{8} (\bibinfo{year}{1966}),
  \bibinfo{pages}{707--710}.
\newblock


\bibitem[Lou et~al\mbox{.}(2019)]%
        {Lou2019}
\bibfield{author}{\bibinfo{person}{Y. Lou}, \bibinfo{person}{J. Chen},
  \bibinfo{person}{L. Zhang}, {and} \bibinfo{person}{D. Hao}.}
  \bibinfo{year}{2019}\natexlab{}.
\newblock \showarticletitle{A survey on regression test-case prioritization}.
\newblock In \bibinfo{booktitle}{\emph{Advances in Computers}}.
  Vol.~\bibinfo{volume}{113}. \bibinfo{pages}{1--46}.
\newblock
\urldef\tempurl%
\url{https://doi.org/10.1016/bs.adcom.2018.10.001}
\showDOI{\tempurl}


\bibitem[Miranda et~al\mbox{.}(2018)]%
        {Miranda2018}
\bibfield{author}{\bibinfo{person}{B. Miranda}, \bibinfo{person}{E. Cruciani},
  \bibinfo{person}{R. Verdecchia}, {and} \bibinfo{person}{A. Bertolino}.}
  \bibinfo{year}{2018}\natexlab{}.
\newblock \showarticletitle{FAST approaches to scalable similarity-based test
  case prioritization}. In \bibinfo{booktitle}{\emph{Proceedings of the
  International Conference on Software Engineering (ICSE)}}.
  \bibinfo{pages}{222--232}.
\newblock
\urldef\tempurl%
\url{https://doi.org/10.1145/3180155.3180210}
\showDOI{\tempurl}


\bibitem[Miranda et~al\mbox{.}(2025)]%
        {FAST2024}
\bibfield{author}{\bibinfo{person}{B. Miranda}, \bibinfo{person}{E. Cruciani},
  \bibinfo{person}{R. Verdecchia}, {and} \bibinfo{person}{A. Bertolino}.}
  \bibinfo{year}{2025}\natexlab{}.
\newblock \bibinfo{title}{FAST replication package}.
\newblock \bibinfo{howpublished}{\url{https://github.com/icse18-FAST/FAST}}.
\newblock
\newblock
\shownote{[Last accessed: 25-March-2025]}.


\bibitem[Noor and Hemmati(2015)]%
        {Noor2015}
\bibfield{author}{\bibinfo{person}{T.~B. Noor} {and} \bibinfo{person}{H.
  Hemmati}.} \bibinfo{year}{2015}\natexlab{}.
\newblock \showarticletitle{A similarity-based approach for test case
  prioritization using historical failure data}. In
  \bibinfo{booktitle}{\emph{Proceedings of the International Symposium on
  Software Reliability Engineering (ISSRE)}}. \bibinfo{pages}{58--68}.
\newblock
\urldef\tempurl%
\url{https://doi.org/10.1109/ISSRE.2015.7381799}
\showDOI{\tempurl}


\bibitem[Pacheco and Ernst(2007)]%
        {Pacheco2007}
\bibfield{author}{\bibinfo{person}{C. Pacheco} {and} \bibinfo{person}{M.~D
  Ernst}.} \bibinfo{year}{2007}\natexlab{}.
\newblock \showarticletitle{Randoop: Feedback-directed random testing for
  Java}. In \bibinfo{booktitle}{\emph{Proceedings of the Companion to the
  SIGPLAN Conference on Object-Oriented Programming Systems and Applications
  Companion (OOPSLA Companion)}}. \bibinfo{pages}{815--816}.
\newblock
\urldef\tempurl%
\url{https://doi.org/10.1145/1297846.1297902}
\showDOI{\tempurl}


\bibitem[Rothermel et~al\mbox{.}(1999)]%
        {Rothermel1999}
\bibfield{author}{\bibinfo{person}{G. Rothermel}, \bibinfo{person}{R.~H.
  Untch}, \bibinfo{person}{C. Chu}, {and} \bibinfo{person}{M.~J. Harrold}.}
  \bibinfo{year}{1999}\natexlab{}.
\newblock \showarticletitle{Test case prioritization: An empirical study}. In
  \bibinfo{booktitle}{\emph{Proceedings of the International Conference on
  Software Maintenance (ICSM)}}. \bibinfo{pages}{179--188}.
\newblock
\urldef\tempurl%
\url{https://doi.org/10.1109/ICSM.1999.792604}
\showDOI{\tempurl}


\bibitem[Rothermel et~al\mbox{.}(2001)]%
        {Rothermel2001}
\bibfield{author}{\bibinfo{person}{G. Rothermel}, \bibinfo{person}{R.~H.
  Untch}, \bibinfo{person}{C. Chu}, {and} \bibinfo{person}{M.~J. Harrold}.}
  \bibinfo{year}{2001}\natexlab{}.
\newblock \showarticletitle{Prioritizing test cases for regression testing}.
\newblock \bibinfo{journal}{\emph{IEEE Transactions on Software Engineering}}
  \bibinfo{volume}{27}, \bibinfo{number}{10} (\bibinfo{year}{2001}),
  \bibinfo{pages}{929--948}.
\newblock
\urldef\tempurl%
\url{https://doi.org/10.1109/32.962562}
\showDOI{\tempurl}


\bibitem[Schott et~al\mbox{.}(2024)]%
        {Schott2024}
\bibfield{author}{\bibinfo{person}{S. Schott}, \bibinfo{person}{S.~E. Ponta},
  \bibinfo{person}{W. Fischer}, \bibinfo{person}{J. Klauke}, {and}
  \bibinfo{person}{E. Bodden}.} \bibinfo{year}{2024}\natexlab{}.
\newblock \showarticletitle{{Java Bytecode Normalization for Code Similarity
  Analysis}}. In \bibinfo{booktitle}{\emph{European Conference on
  Object-Oriented Programming (ECOOP)}}, Vol.~\bibinfo{volume}{313}.
  \bibinfo{pages}{37:1--37:29}.
\newblock
\urldef\tempurl%
\url{https://doi.org/10.4230/LIPIcs.ECOOP.2024.37}
\showDOI{\tempurl}


\bibitem[Srikanth et~al\mbox{.}(2005)]%
        {Srikanth2005}
\bibfield{author}{\bibinfo{person}{H. Srikanth}, \bibinfo{person}{L. Williams},
  {and} \bibinfo{person}{J. Osborne}.} \bibinfo{year}{2005}\natexlab{}.
\newblock \showarticletitle{System test case prioritization of new and
  regression test cases}. In \bibinfo{booktitle}{\emph{International Symposium
  on Empirical Software Engineering}}. \bibinfo{pages}{10--pp}.
\newblock
\urldef\tempurl%
\url{https://doi.org/10.1109/ISESE.2005.1541815}
\showDOI{\tempurl}


\bibitem[Vargha and Delaney(2000)]%
        {Vargha2000}
\bibfield{author}{\bibinfo{person}{A. Vargha} {and} \bibinfo{person}{H.~D
  Delaney}.} \bibinfo{year}{2000}\natexlab{}.
\newblock \showarticletitle{A critique and improvement of the CL common
  language effect size statistics of McGraw and Wong}.
\newblock \bibinfo{journal}{\emph{Journal of Educational and Behavioral
  Statistics}} \bibinfo{volume}{25}, \bibinfo{number}{2}
  (\bibinfo{year}{2000}), \bibinfo{pages}{101--132}.
\newblock
\urldef\tempurl%
\url{https://doi.org/10.3102/10769986025002101}
\showDOI{\tempurl}


\bibitem[Wang et~al\mbox{.}(2015)]%
        {Wang2015}
\bibfield{author}{\bibinfo{person}{R. Wang}, \bibinfo{person}{S. Jiang}, {and}
  \bibinfo{person}{D. Chen}.} \bibinfo{year}{2015}\natexlab{}.
\newblock \showarticletitle{Similarity-based regression test case
  prioritization}. In \bibinfo{booktitle}{\emph{Proceedings of the
  International Conference on Software Engineering and Knowledge Engineering
  (SEKE)}}. \bibinfo{pages}{358--363}.
\newblock
\urldef\tempurl%
\url{https://doi.org/10.18293/seke2015-115}
\showDOI{\tempurl}


\bibitem[Wu et~al\mbox{.}(2012)]%
        {Wu2012}
\bibfield{author}{\bibinfo{person}{K. Wu}, \bibinfo{person}{C. Fang},
  \bibinfo{person}{Z. Chen}, {and} \bibinfo{person}{Z. Zhao}.}
  \bibinfo{year}{2012}\natexlab{}.
\newblock \showarticletitle{Test case prioritization incorporating ordered
  sequence of program elements}. In \bibinfo{booktitle}{\emph{Proceedings of
  the International Workshop on Automation of Software Test (AST)}}.
  \bibinfo{pages}{124--130}.
\newblock
\urldef\tempurl%
\url{https://doi.org/10.1109/IWAST.2012.6228980}
\showDOI{\tempurl}


\bibitem[Yoo and Harman(2010)]%
        {Yoo2010B}
\bibfield{author}{\bibinfo{person}{S. Yoo} {and} \bibinfo{person}{M. Harman}.}
  \bibinfo{year}{2010}\natexlab{}.
\newblock \showarticletitle{Regression testing minimization, selection and
  prioritization: A survey}.
\newblock \bibinfo{journal}{\emph{Software testing, verification and
  reliability}} \bibinfo{volume}{22}, \bibinfo{number}{2}
  (\bibinfo{year}{2010}), \bibinfo{pages}{67--120}.
\newblock
\urldef\tempurl%
\url{https://doi.org/10.1002/stvr.430}
\showDOI{\tempurl}


\end{thebibliography}

\end{document}